\documentclass[%
 aip,
 amsmath,amssymb,
 reprint,%
]{revtex4-1}

\usepackage{breqn}
\usepackage{graphicx}
\usepackage{dcolumn}
\usepackage{bm}

\usepackage[utf8]{inputenc}
\usepackage[T1]{fontenc}
\usepackage{mathptmx}
\usepackage{etoolbox}

\makeatletter
\def\@email#1#2{%
 \endgroup
 \patchcmd{\titleblock@produce}
  {\frontmatter@RRAPformat}
  {\frontmatter@RRAPformat{\produce@RRAP{*#1\href{mailto:#2}{#2}}}\frontmatter@RRAPformat}
  {}{}
}%

\usepackage{graphicx}

\makeatother
\begin{document}

\preprint{AIP/123-QED}

\title[Realization of an advanced super-mirror solid-state neutron polarizer for the instrument PF1B at the Institut Laue-Langevin]{Realization of an advanced super-mirror solid-state neutron polarizer for the instrument PF1B at the Institut Laue-Langevin}
\author{A.K. Petoukhov}
\author{V.V. Nesvizhevsky}
\author{T. Bigault}
\author{P. Courtois}
\author{A. Devishvili}
\author{D. Jullien}
\author{T. Soldner}
\email{petukhov@ill.fr, nesvizhevsky@ill.eu}
\affiliation{Institut Max von Laue - Paul Langevin$,$ 71 avenue des Martyrs$,$ 38042 Grenoble$,$ France}

\date{\today}

\begin{abstract}
In this last of a series of three papers on the development of an advanced solid-state neutron polarizer, we present the final construction of the polarizer and the results of its commissioning. The polarizer uses spin-selective reflection of neutrons by interfaces coated with polarizing super-mirrors. The polarizer is built entirely in-house for the PF1B cold neutron beam facility at the Institut Max von Laue - Paul Langevin (ILL). It has been installed in the PF1B casemate and tested at real conditions. The average transmission for the ``good'' spin component is measured to be >30~\%. The polarization averaged over the capture spectrum reaches a record value of $P_n\approx 0.997$ for the full angular divergence in the neutron beam, delivered by the H113 neutron guide, and the full wavelength band $\lambda$ of $0.3-2.0~nm$. This unprecedented performance is due to a series of innovations in the design and fabrication in the following domains: choice of the substrate material, super-mirror and anti-reflecting multilayer coatings, magnetizing field, assembling process. The polarizer is used for user experiments at PF1B since the last reactor cycle in 2020. 
\end{abstract}

\maketitle

\section{Introduction}

PF1B is a user facility at the Institut Max von Laue - Paul Langevin (ILL) in Grenoble, France, for experiments in elementary particle and nuclear physics using polarized or non-polarized cold neutrons. PF1B is located at the end position of the ``ballistic'' $m=2$ super-mirror cold neutron guide H113 with an exit cross section of $60\times 200~mm^2$ ~\cite{Abe2006nima} (note that the capture flux at the guide exit of $1.35\times 10^{10}~n/cm^2/s$ reported in~\cite{Abe2006nima} has been improved to $2.2\times 10^{10}~n/cm^2/s$ at nominal reactor power by replacing guide sections that had suffered radiation damage and by upgrading the in-pile part). 

An important component of PF1B is a cold neutron polarizer that has to produce a large-area, $\sim 80\times 80~mm^2$, well-polarized neutron beam over an extended range of neutron wavelengths, $0.3-2.0~nm$. For the needs of experiments of different types, PF1B has to provide several options of optimization including a maximum total flux of polarized neutrons over a large beam cross section, a maximum flux density of polarized neutrons over a relatively small beam cross section, and ultra-high precision of the knowledge of the polarized neutron beam properties. While in the first two cases, the average polarization can be moderate (typically $P_n > 0.98$), the latter option requires ultra-high polarization ($P_n \geq 0.997$) to minimize systematic uncertainties ~\cite{Ves2008prc,Gle2017plb,Goe2007plb,Gag2016prc,Mar2019prl,Kre2005plb}. 

In order to achieve high polarization levels with reasonable transmission over the full wavelength range, the preferred technology is typically super-mirror (SM) benders ~\cite{Mez1977cp,Dra1977jtp,Sch1989pb,Maj1995pb,Mar2007tsf,Kri2008,Mez1989spie}. Ultra-high polarization is difficult to achieve with single reflection or transmission by a polarizing SM, though noticeable investigation was made in that direction \cite{Ple2010nima}. Therefore, the design of the polarizer geometry aims at making most neutrons reflect at least twice on polarizing SMs, with sufficiently large angle of incidence, when going through the device \cite{Kre2005nima,Pet2016nima}.

The previous PF1B polarizer was built using the traditional technology of air-gaped reflection-type polarizing benders. It consisted of $30$ channels of $80~cm$ length and $2~mm$ width. The thickness of the borofloat glass substrates was $0.7~mm$. The Co/Ti/Gd SM coatings ~\cite{Sch1989pb,Els1994tsf,Cou2013ILL} had the effective critical velocity of $m=2.8$. The polarizer cross section was $80\times 80~mm^2$, the radius $300~m$, and the applied magnetic field $120~mT$ ~\cite{Sol2002ILL}. This polarizer was produced in collaboration between the ILL (SM coatings) and the TU M{\"u}nchen (glass and assembly). It was installed in 2002 downstream the H113 guide in an effective neutron capture flux of $\sim 2\times 10^{10}~n/cm^2/s$. The measured transmission was $\sim 0.49$ for the ``good'' polarization component. The capture-flux-averaged polarization was $P_n\sim 0.985$. 

When ultra-high polarization was required, we installed a second polarizer of the same type in the ``X-SM geometry'', thus providing a mean polarization of $P_n =0.997$ and a transmission of $\sim 0.25$ for the ``good'' polarization component ~\cite{Kre2005nima}. This geometry assumes that the reflecting planes in the second polarizer are orthogonal to the reflecting planes in the first one. 

However, during more than 15 years of successful exploitation, this polarizer was irradiated with a very high neutron fluence which resulted in a significant irradiation damage to the mirrors' borofloat glass substrates, mainly by the charged particles from the reaction $^{10}B(n,\alpha)$ in the glass substrate. It is also strongly activated, mainly due to the presence of Co in the SM coatings, which makes its handling seriously complicated. 

In this paper, we present the new, Advanced polarizer built for the PF1B instrument at ILL with improved polarization and free from the radiation damage and activation issues. The polarizer design, the choice of substrate and SM coating, and the magnetic housing are described in great details in our previous publications ~\cite{Pet2016nima,Pet2019rsi}. Here, we focus on the full scale polarizer production and the results of measurements of its characteristics.   

\section{\label{sec:Design}Polarizer design}

To avoid some drawbacks of the previous polarizer, namely the high activation of Co in the Co/Ti SM coatings and the neutron-induced degradation associated with the $^{10}B(n,\alpha)$ reaction in the borofloat glass substrates, we decided to build a solid-state polarizer with Fe/Si coatings ~\cite{Maj1995pb,Kri1998pb,Hog1999pb,Stu2006pb,Big2009ILL,Wil2011itr}.

An immediate advantage of the solid-state polarizer is its compactness. It has a more favorable ratio of channel to inter-channel width and allows to design a magnetic system with better performance. 

In traditional C-benders, a solid-state polarizer is built of a bent stack of thin ($150-200~\mu m$) single crystal Si wafers, coated on both sides with Fe/Si SM coatings terminated by Gd absorbing layers. Each Si plate coated with two reflecting SMs is a spin-dependent guide for neutrons which enter the plate bulk through the entrance edge of the plate. To avoid the direct view (i.e.\ neutron trajectories which do not touch the polarizing coatings) the bending angle $\gamma_C$ should meet the following condition:

\begin{equation}
\gamma_C\geq 8d/L,
\label{eq:gammac}
\end{equation}
where $d$ and $L$ are thickness and full length of the channel. A double-bent polarizer of this type is known as the S-bender~\cite{Stu2006pb,Big2009ILL,Wil2011itr}. 

In our design of the new PF1B polarizer, we follow the concept proposed and described in detail in our previous publications~\cite{Pet2016nima,Pet2019rsi}. According to this ``advanced'' concept, we replace the single-crystal Si substrates by single-crystal sapphire. Since the neutron-optical potential for sapphire is higher than that for spin-down neutrons in the magnetized Fe of the SM coatings, this choice allows us to avoid the total reflection regime for neutrons of the unwanted spin direction propagating through the substrates. This modification expands the efficient polarizer bandwidth into the low $Q$ region. 

The polarizer is built of two independent stacks of flat substrate plates of $80\times 25\times 0.18~mm^3$ (i.e.\ $L/2=25~mm$, $d=0.18~mm$), each coated with SMs on both sides. Each plate in the stack is mounted on the top of the previous one. The total number of plates in the stack is 440 thus providing the total polarizer cross section of $80\times 80~mm^2$. 

The two stacks of mirrors are tilted by angle $\gamma_\nu$:
\begin{equation}
\gamma_\nu\geq 4d/L.
\label{eq:gammanu}
\end{equation}

We denote this type of polarizer the V-bender.

\begin{figure}
\centering
\includegraphics[width=\columnwidth]{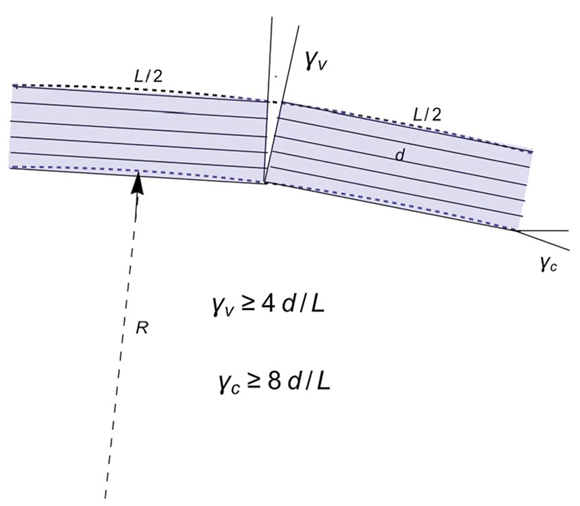}
\caption{\label{fig:Vbender} The V-bender geometry: two stacks of both-side SM-coated plane parallel substrate plates (with thickness $d$ and length $L/2$) are tilted by angle $\gamma_\nu$ with respect to each other. To avoid the ``direct view'', the angle has to meet condition Eq.~(\ref{eq:gammanu}). Note that the two halves of the V-bender do not need to be in contact, also the slits do not need to be lined up. The traditional C-bender geometry is indicated by dashed lines for comparison (the internal dashed lines are omitted for better readability).}
\end{figure}

Fig.~\ref{fig:Vbender} illustrates the V-bender geometry in comparison with the traditional C-bender geometry.

An important feature of the V-bender is the absence of pronounced Bragg dips in the transmission for particular wavelength values. Such dips were observed in experiments with solid-state S- and C-benders when the angular divergence of the incident neutron beam is comparable to the bending angle~\cite{Stu2006pb,Sha2014nima}. For the reflection of neutrons from a flat perfect crystal, the acceptance angle of Bragg reflection is very small (typically $1-10~\mu rad$) and the corresponding dip would be completely washed out by the angular divergence of the incident beam (typically a few tens $mrad$) even in the case of an unfortunately chosen crystal orientation. Indeed, these expectations were confirmed in experiment as we observed no dip in the transmission. 

As mentioned in ~\cite{Kat2018pcgcm,Pet2019rsi}, sapphire substrates are readily available ~\cite{Siegert} with sufficient polishing quality for SM coating and minimize the substrate bending due to the residual stress present in the coating. The latter is expected to be relevant when considering the geometrical imperfections of the final assembled mirror stack with respect to the ideal one. Together with the neutron optical properties, these features led us to choose sapphire as substrate material.

\section{Polarizing coating}

To explore the full angular divergence of the H113 ballistic $m=2$ SM guide and to cover the broad wavelength band of $0.3-2.0~nm$, we use the same ``inverse'' scheme $m=3.2$ Fe/SiNx SM coatings, consisting of 603 individual layers, which was previously used for the production of a solid-state S-bender~\cite{Stu2006pb,Big2009ILL}. The term ``inverse'' refers to the sequence order for the SM coating depositions, starting with a thicker layer as opposed to the sequence in SMs used for neutrons incident from air. Note that with such a sequence in a solid-state polarizer, the first layer visible to neutrons is the thickest one, as in the case of air-gaped devices. 

An absorbing Gd layer also has to be coated on top of the SM so that neutrons which are transmitted through the SM do not get out of the polarizer or into the next plate. In order to guarantee that these neutrons are absorbed rather than reflected, even at low $Q$ values, an anti-reflecting and absorbing Si/Gd multilayer consisting of 41 individual layers and based on the same principle as in Ref.~\cite{Sch1994pb} was designed. The total thickness of Gd is larger than $500~nm$, assuring that the transmission of non-reflected neutrons through the interface between the plates stacked inside the polarizer is always well bellow $10^{-3}$ in the operation conditions.

The SM coating was produced in-house by reactive magnetron sputtering~\cite{Hog1999pb,Big2014jpcs}. Since the S-bender production~\cite{Stu2006pb,Big2009ILL}, some investigations were made about the magnetic properties of the coatings~\cite{Mar2016nima,Mar2019cry} and the neutron beam depolarization at reflection~\cite{Kla2013pp,Kla2016nima}. The coating process had been optimized further, resulting in magnetically softer multi-layers, i.e.\ requiring a weaker applied magnetic field to be magnetized close to saturation. 

\begin{figure}
\centering
\includegraphics[width=\columnwidth]{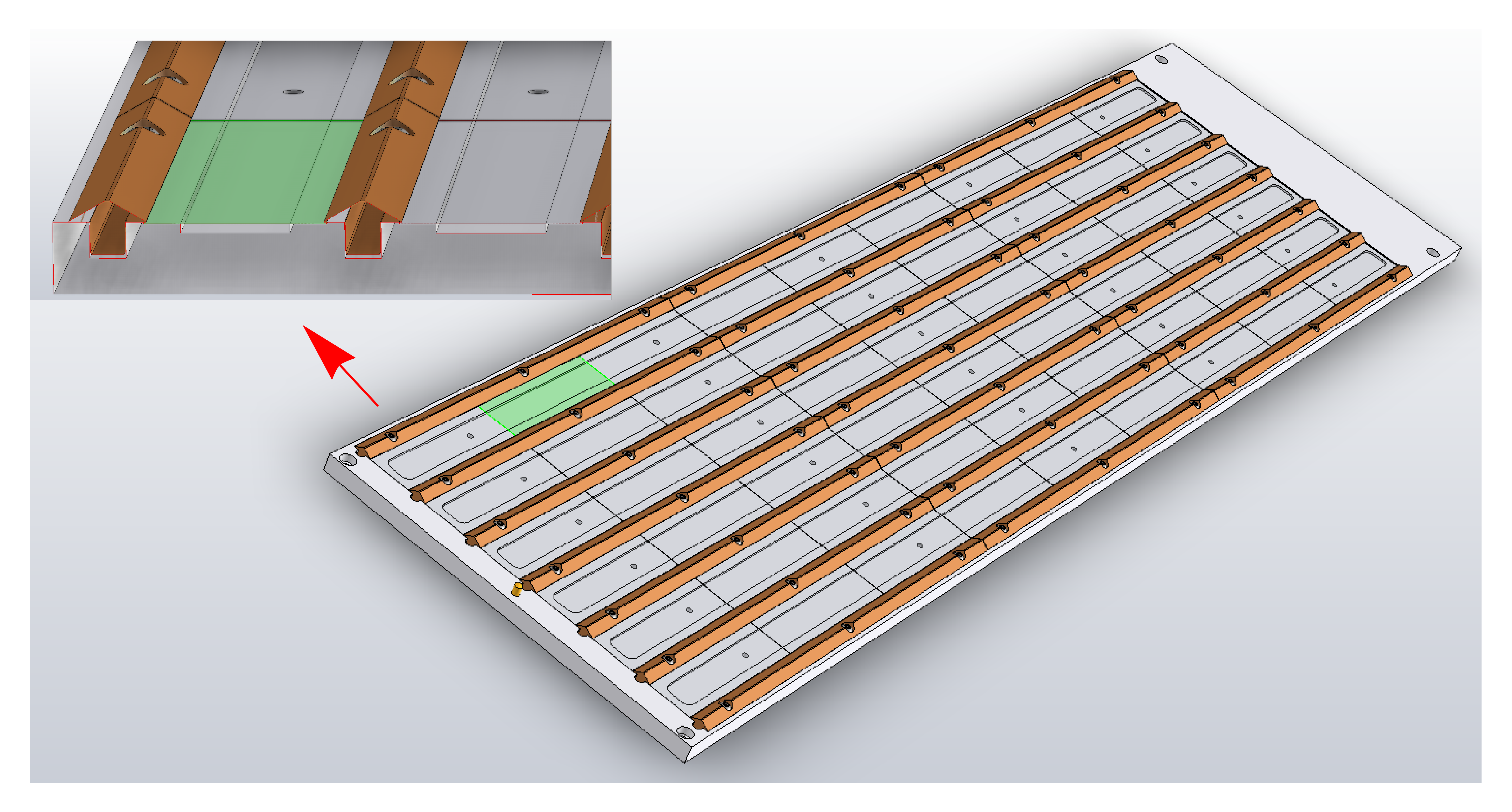}
\caption{\label{fig:Plateau} Scheme of the deposition tray. The production version contains 48 sapphire plates of dimension $80\times25\times 0.18~mm^3$ each, a witness float glass sample and a witness thick sapphire plate with dimensions $80\times40\times 3~mm^3$ for neutron reflectometry measurements. These 2 additional substrates were placed in the free space (top-right corner). One thin sapphire substrate is highlighted in green, and the inset shows a detail of the triangular-shaped separator, used for masking the mirror edges without reducing the thickness deposited on their faces.}
\end{figure}

With our in-line sputtering machines, we used the tray shown in Fig.~\ref{fig:Plateau} to coat one face on a set of 48 sapphire substrates at once. In order to prevent coating the edges of the plates (where in particular Gd would reduce the neutron transmission significantly), special care was taken in designing the deposition tray, so that each mirror face is maximally coated without depositing material on the edges. In practice, an area of about $0.3~mm$ wide along the mirror edges was masked for this purpose. A witness float glass sample and a witness sapphire sample of $80\times40\times 3~mm^3$ were coated together with each set. All single-crystal sapphire substrates ($0.18~mm$ plates and $3~mm$ witness plates), with the c-plane parallel to the surface, were from ~\cite{Siegert}. For technical reasons, each coating was made in two steps with two different machines: one for the SM and one for the anti-reflecting absorbing multi-layer. The typical production cycle, for coating a set of 48 plates on both sides, was about one week. Twenty-five such cycles were achieved, spanning about six months, resulting in a total SM coated area of about $4~m^2$. 

For each coating run, the witness float glass sample was removed from the tray after the SM coating, so that it can be measured by neutron reflectometry in the standard way, with neutrons coming from the ``air'' side. The witness sapphire samples of $3~mm$ thickness underwent both steps and were measured with neutrons entering by the substrate edge, reflecting at the SM from the substrate side. This allowed a systematic control of each coating performance in the same conditions as for the thin plates, which could not be measured by neutron reflectometry. 

\begin{figure}
\centering
\includegraphics[width=\columnwidth]{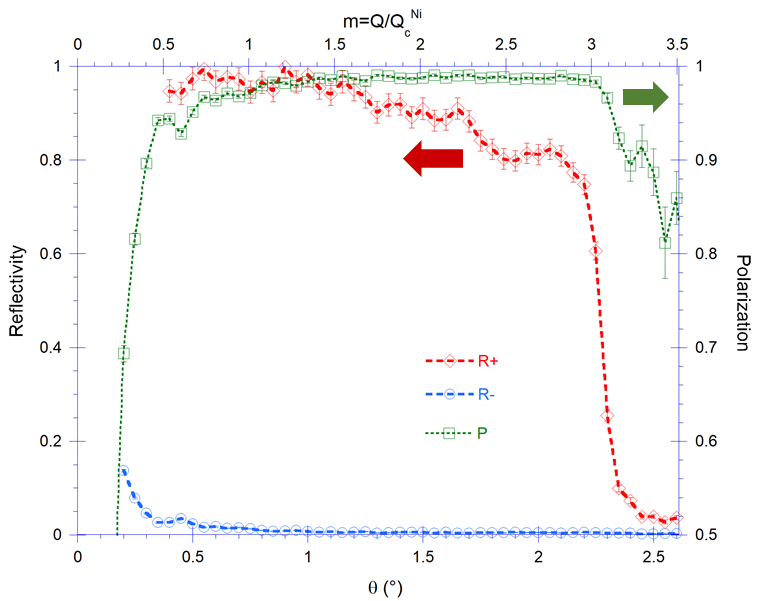}
\caption{\label{fig:Reflectivity} Spin-dependent reflectivity ($R_{+}=R_{++}+R_{-+}$: red, $R_{-}=R_{--}+R_{+-}$: blue; left axis) measured with the T3 instrument on a $3~mm$ thick sapphire witness sample coated with Fe/SiNx/Gd SM and anti-reflecting absorbing layer in a production run. Neutrons with wavelength $0.75~nm$ entered by the substrate edge and were reflected at the SM from the sapphire side. The polarization (P: green; right axis) was calculated from $R_{+}$ and $R_{-}$, applying only a background correction.}
\end{figure}

Fig.~\ref{fig:Reflectivity} shows typical spin-dependent reflectivity $R_{+},R_{-}$ measured with our test reflectometer T3 for one of the $3~mm$-thick sapphire witness samples. Most of the features, in particular the low-$Q$ part of the ``$R_{-}$'' curve, are consistent with the simulations presented in ~\cite{Pet2019rsi}. Through the whole production, the measured polarization after single reflection was $>0.985$ in the range $1<m<3$. 

\begin{figure}
\centering
\includegraphics[width=\columnwidth]{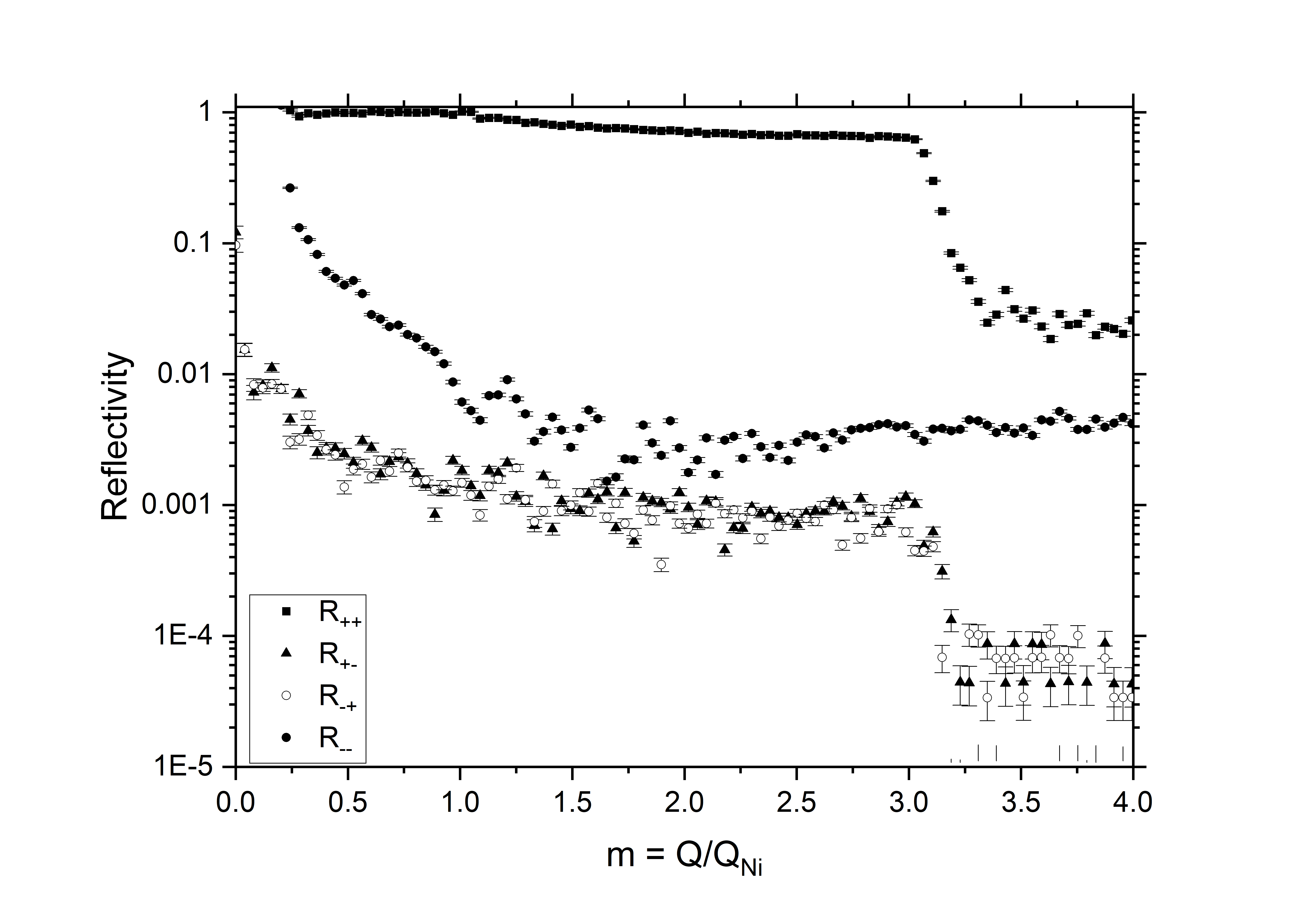}
\caption{\label{fig:RvsQSAdam} Spin-dependent reflectivity $R_{++},R_{+-},R_{-+},R_{--}$ measured at SuperADAM reflectometer with an opaque polarized $^3$He analyzer (the analyzing power of $>0.999$) and the applied magnetic field of $0.7~T$.}
\end{figure}

We also performed more accurate spin-dependent reflectometry measurements at SuperADAM~\cite{Dev2013rsi} equipped with an opaque polarized $^3$He analyzer. Fig.~\ref{fig:RvsQSAdam} shows the results of measurements for all 4 spin components $R_{++},R{+-}, R_{-+},R_{--}$.

 Fig. ~\ref{fig:RvsQSAdam} shows that the beam polarization after single reflection ~\cite{Pet2019rsi} is $0.995<P<0.999$ in the same range $1<m<3$, i.e. it is significantly higher than the value $0.985$ measured at T3. We explain this difference by a lower polarization of the incident neutron beam and imperfections of the spin-flipper at T3 instrument.

\begin{figure}
\centering
\includegraphics[width=\columnwidth]{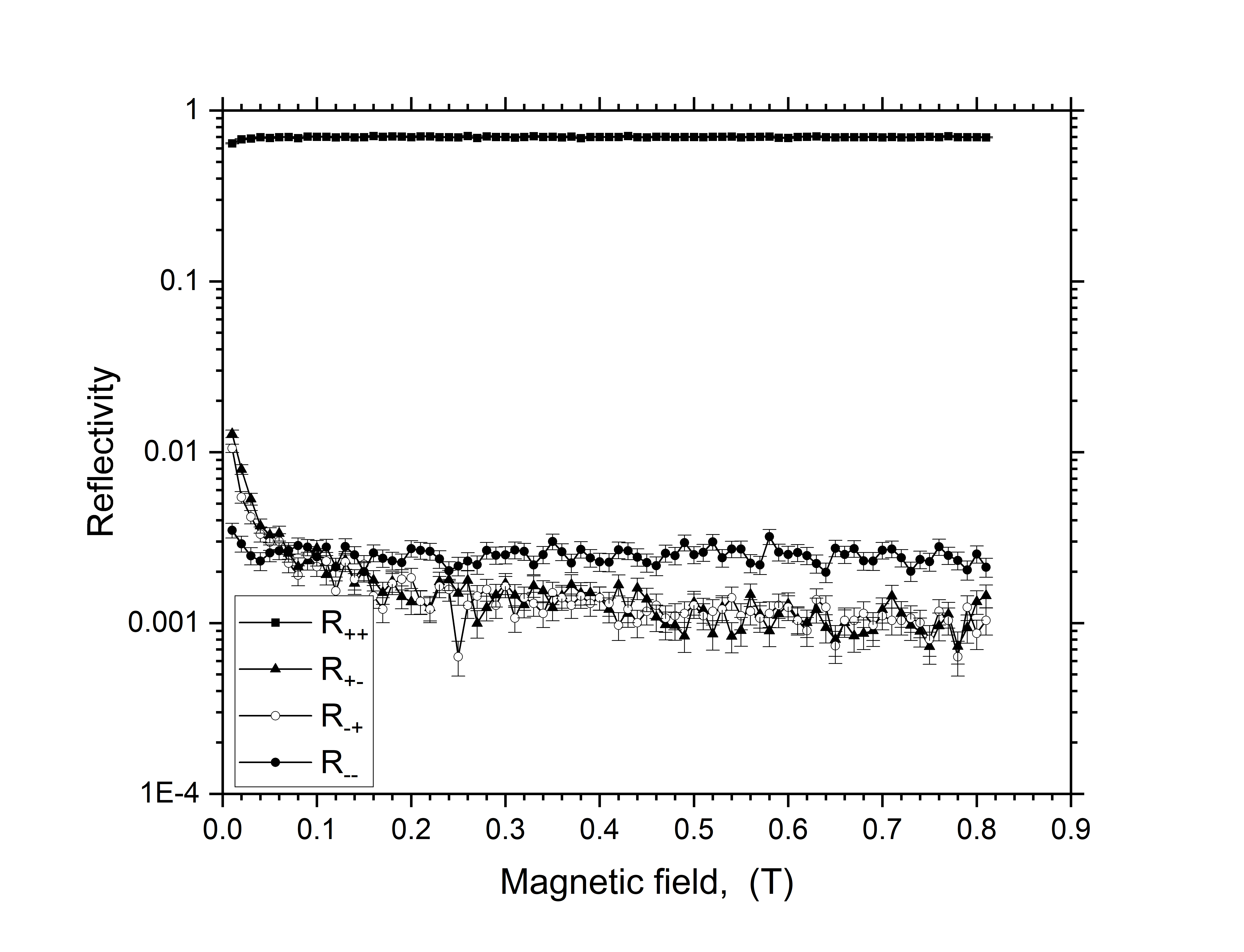}
\caption{\label{fig:RvsBSAdam} Spin-dependent reflectivity $R_{++},R_{+-},R_{-+},R_{--}$ versus the applied magnetic field measured at SuperADAM with an opaque polarized $^3$He analyzer (the analyzing power of $>0.999$). $m=1.92$.}
\end{figure}

For a polarizer based on multiple reflections, the resulting polarization is defined by the depolarization at the last reflection ~\cite{Pet2019rsi}. Therefore, we also measured at SuperADAM the spin-dependent reflectivity as a function of the applied magnetic field strength, see FIG. ~\ref{fig:RvsBSAdam}. In contrast to $R_{++}$ and $R_{--}$ that are field-independent for $B>10~mT$, the reflectivity $R_{+-}$ and $R_{-+}$ decays rapidly at the field strengths of up to $100~mT$, and continues decaying slowly at larger fields. This confirms the importance of a high magnetizing field and justifies our magnetic housing with $0.38~T$ permanent magnet ~\cite{Pet2019rsi} for the PF1B polarizer.   

\section{Assembling}

\subsection{\label{sec:Control precision}In-situ procedure to control the assembling precision}

In solid-state neutron polarizers, the plates with the polarizing mirrors are mounted on the top of each other thus forming a cassette which is needed in order to cover a significant cross section of the neutron beam. It is usually assumed that all plates are nearly identical and have a plane-parallel geometry. 

Real plates can differ from plane-parallel as a result of the polishing process and due to residual stress in the coating. Imperfections of the geometry of individual plates as well as dust particles between them would result in a scatter of individual angles between the reflecting surfaces and the incident beam direction. The errors in setting the inclination angle would result in neutron reflection losses and an increased angular divergence of the reflected beam. The width of such dispersion increases with the number of plates in the cassette according to the law of ``Gaussian Random Walk'', see Appendix~\ref{app:RestrictionOnMirrorNumber}. 

The primary effect of such an angular dispersion is the degradation of the polarizer transmission for the ``good'' spin component, provided that the width of the dispersion is comparable to the critical angle of the polarizer (typically $10-20~mrad$). This mechanism may explain the significant discrepancy between the expected and measured transmission often observed in experiments with solid-state neutron-optical devices ~\cite{Sha2014nima}. Some polarization loss may also be attributed to this effect since such an angular spread may make neutron trajectories without reflection possible. 

The most straightforward way to minimize the losses of efficiency due to this mechanism would be to improve the precision in setting individual plates and to reduce the number of plates in the cassette. However, the number of plates in the cassette is fixed by the size of the required beam, and setting the inclination angle of individual plates with the precision much better than $\pm1~mrad$ is challenging.

\begin{figure}
\centering
\includegraphics[width=\columnwidth]{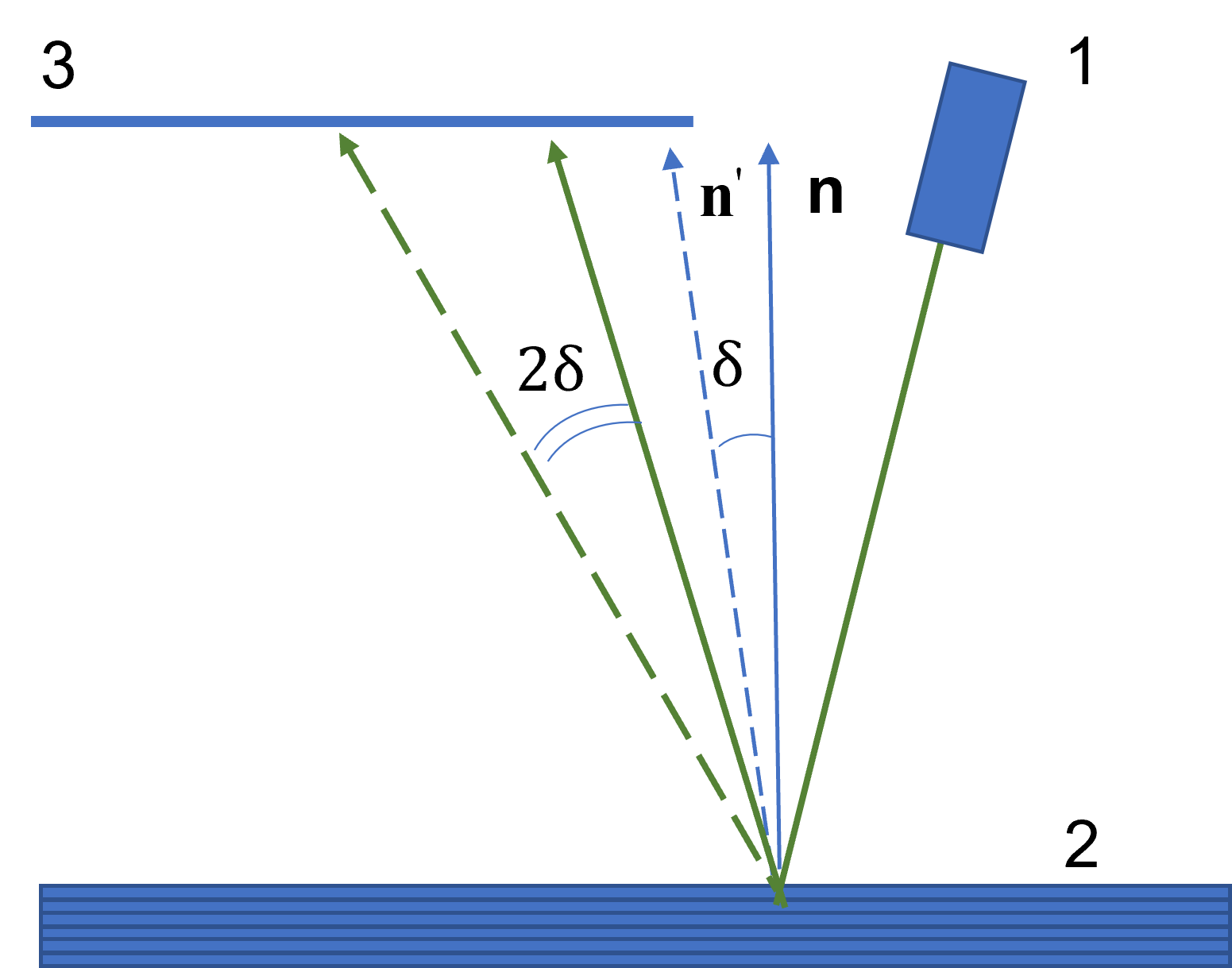}
\caption{\label{fig:Scheme} Scheme of the in-house-built optical inclinometer: 1 - laser, 2 - plates in the stack, 3 - full-frame sensor of a digital photo-camera.}
\end{figure}

An advantage of the rather simple V-bender geometry is the possibility to develop a procedure to actively control the inclination angle of each plate in the cassette, see Fig.~\ref{fig:Scheme}. The idea is to limit the cumulative effect of successive random plate misalignment, by making it ``less'' random through deterministic control and intervention on each plate. The stack of assembled plates is illuminated with a narrow laser beam (diameter $1~mm$), and the reflected beam is projected on a full-frame ($24\times 36~mm$) sensor of a digital photo-camera. Assuming $\boldsymbol{n}$ is normal to the top mirror of the reference plate and $\boldsymbol{n'}$ is normal to the top mirror of the inspected plate, which is inclined relative to $\boldsymbol{n}$ by a small angle $\delta$, the shift $\Delta$ of the spot position on the camera sensor is:
\begin{equation}
\Delta \approx 2\delta~h,
\label{eq:Delta}
\end{equation}
where $h\approx1000~mm$ is the distance between the plates and the sensor. We take the position of the reflected spot from the very first plate as the reference. If we observe that $\Delta$ is outside the accepted tolerance (typically $\pm~1~mrad$), we carefully inspect this plate and usually find a large dust particle. After removing it, the plate is assembled back to the cassette and we continue the procedure.

We found that at normal laboratory conditions, $\sim 0.25$ of the plates show an angular deviation $>1~mrad$. Therefore, we decided to perform all manipulations with the sapphire plates (before and after SM deposition) inside our class-100 laminar-flow box. This solution dramatically improved the output of ``good'' plates: practically all of them show a small spread of the inclination angle ($<1~mrad$). During the stacking procedure, when misalignment with the underlying plate of $>1~mrad$ occurs, there is still a possibility to flip the last plate upside down and check if the misalignment is reduced. In the rare case when this did not solve the problem, the last plate was replaced by another one. 

\begin{figure}
\centering
\includegraphics[width=\columnwidth]{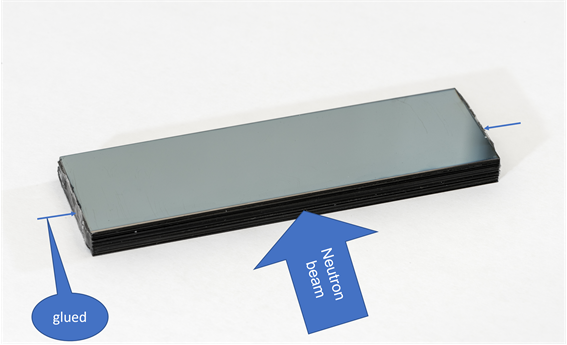}
\caption{\label{fig:InterStack} An intermediate stack of 25 sapphire plates. The neutron beam is incident on the long side ($80~mm$). The position of each individual plate is laser controlled according to the scheme shown in Fig.~\ref{fig:Scheme}. Ultraviolet cured glue is applied to both opposite short sides ($25~mm$).}
\end{figure}

To simplify the polarizer assembling we decided to use intermediate stacks composed of 25-30 plates. During assembling, the orientation of each plate is laser controlled according to the scheme shown in Fig.~\ref{fig:Scheme} and, finally, the positions of all plates in the intermediate stack are fixed with UV-cured optical glue NOA65 ~\cite{Tho} applied to both opposite short sides, see Fig.~\ref{fig:InterStack}.

\begin{figure}
\centering
\includegraphics[width=\columnwidth]{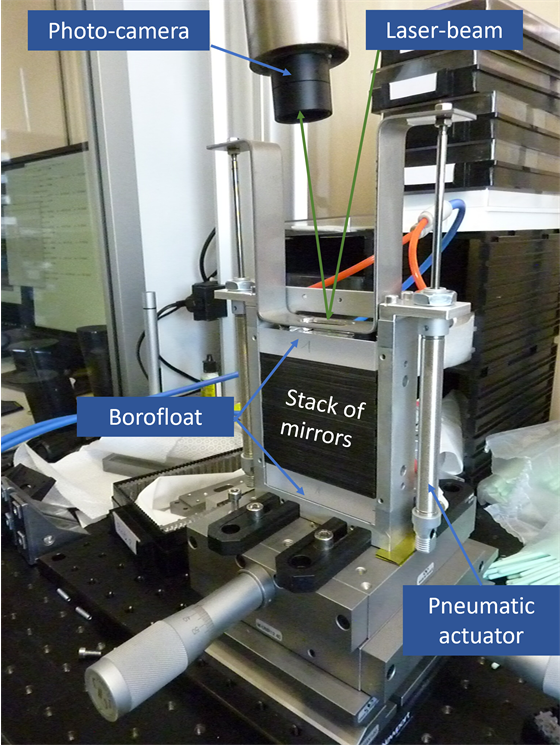}
\caption{\label{fig:Assembling} Final assembling of the polarizer cassette. Intermediate stacks are confined between two thick ($25~mm$) Borofloat glass flats coated with Al. A green-line laser beam incident on the upper flat and the reflected beam are registered with a photo-camera. Two pneumatic actuators apply homogeneous pressure ($\sim 1~bar$) on the cassette.}
\end{figure}

Further assembling of polarizing cassettes composed of intermediate stacks also was performed in the same ``clean room'' with the optical control and under an applied load, see Fig.~\ref{fig:Assembling}. Intermediate stacks were inserted between two optically polished flats made of Borofloat glass and mounted on top of each other. The Borofloat flats and the cassette body made from a B-Al compound serve also as a neutron diaphragm absorbing neutrons outside the polarizer aperture. Two pneumatic actuators apply homogeneous load to the cassette ($\sim 1~bar$) in order to minimize possible gaps between the plates. The fully assembled cassette of plates is fixed with non-magnetic screws.

\subsection{\label{sec:Neutron test}Neutron test of the assembly accuracy}

We may suspect that the optical inspection of the assembling described above is valid only locally, within the laser spot size of $\sim 1~mm$. The global slope (averaged over the full plate surface) may be different. Therefore, we also performed neutron tests of the assembly quality at the Super-Adam reflectometer at the ILL ~\cite{Dev2013rsi}.  

A fully assembled cassette of 440 double-sided coated plates was installed in the target position of the instrument. A very narrow neutron beam (width $0.1~mm$, horizontal angular divergence $0.05~mrad$ FWHM, height $60~mm$) with the wavelength of $\lambda =0.5~nm$ was incident on the cassette. After preliminary alignment, the cassette was tilted by an angle of $\sim 7~mrad$ from the incident beam direction. This angle is close to the value $\theta_d=d/L=7.2~mrad$ needed to avoid direct transmission, although a small admixture of neutrons having experienced double and zero reflections is possible due to the geometrical imperfections of the plates. 

\begin{figure*}
\centering
\includegraphics[width=\textwidth]{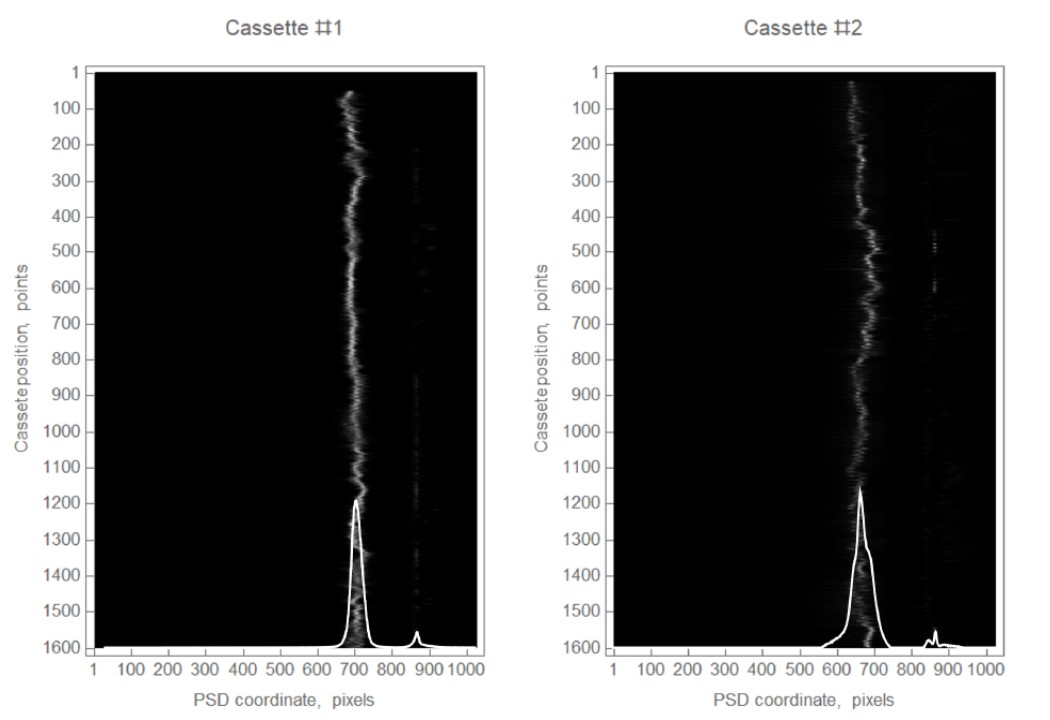}
\caption{\label{fig:CassetteScan} Two-dimensional matrix representing the angular distribution of neutrons reflected by the mirrors of the 440 individual plates in Cassette \#1 (Left) and Cassette \#2 (Right). The solid white line shows the projection on the horizontal axis. See text for details.}
\end{figure*}

The angular distribution of neutrons having passed through the cassette is projected on a position-sensitive detector (PSD) with a position resolution of $2.8~mm$ FWHM, installed at distance $r$ from the cassette (Cassette \#1: $r=2500~mm$, Cassette \#2: $r=3250~mm$). During the experiment, we keep the incident beam position and the PSD position, while the cassette position is scanned across the beam with a step of $0.05~mm$. The full range of the cassette position scan is $440\times 0.18~mm = 80~mm$. The data were accumulated in two-dimensional matrices: cassette position versus PSD pixel number, see Fig.~\ref{fig:CassetteScan}. 

Data for Cassette \#1 were re-binned to the same sample-detector distance $r\approx 3250~mm$ as for Cassette \#2. The bright spots near PSD pixel $\sim 700$ are from neutrons having experienced a single reflection by the mirrors. Much weaker spots near PSD pixel $\sim 870$ are from neutrons having experienced zero reflection. As it is common in reflectometry, the distance $\Delta$ between the position of zero reflection and the single reflection spot is related to the mirror slope $\theta$ as follows:
\begin{equation}
\Delta \approx 2\theta r.
\label{eq:Delta2}
\end{equation}

Note that for the parameters of our experiment, the positions of the zero-reflection spots are independent of the mirror slope, and the width of the spot is dominated by the PSD's finite resolution, $\sigma_{\rm PSD}\approx 4.25$ pixels, or $\sim 0.37~mrad$. 

In contrast, the positions of the single reflection spot are defined by the angles of the individual mirrors. 

Since the incident beam width is $0.1~mm$, the cassette position step $0.05~mm$, and the plate thickness $\sim 180~\mu m$, the beam often illuminates two adjacent plates simultaneously resulting in an additional broadening of the spot or even a spot splitting due to an eventual difference of slopes. At the same time, the width of the narrowest spots is driven by the PSD resolution and the mirror waviness.

We observed that due to the mechanical polishing our sapphire substrates are typically thinner near the edges than they are in the central region. This difference in the thickness is $\sim 6~\mu m$. Reflection of neutrons from a concave surface seen during their propagation through the plate would result in focusing in space and in the defocusing (additional broadening) in angle.

From the data shown in Fig.~\ref{fig:CassetteScan}, we found that the typical width of the most narrow spots (the beam illuminates a single mirror) is very close to $\sigma_{PSD}$ that allows us to conclude that the individual mirror waviness of our sapphire substrates play a minor role in the formation of the reflected beam angular distribution.

By projecting the matrix data on the horizontal axis we obtain the effective angular distribution of neutrons reflected by the mirrors of all 440 plates. These angular distributions measured for both cassettes are shown in Fig.~\ref{fig:CassetteScan} with solid white lines. 

By fitting the single reflection peak near pixel $\sim 700$ with a Gaussian we obtain the estimates of standard deviations $\sigma_\theta^1$ and $\sigma_\theta^2$ for the distribution of individual slopes in Cassettes \#1 and \#2:
\begin{equation}
\sigma_\theta^1=0.62 ~ mrad,\quad \sigma_\theta^2=1.02 ~ mrad.
\label{eq:sigmas}
\end{equation}

As follows from Eq.~(\ref{eq:sigmas}), Cassette \#2 shows a significantly broader distribution of mirror slopes. This fact may be explained by the difference in the assembling procedure. Indeed, Cassette \#1 was assembled from intermediate stacks in a single run. In contrast, while mounting Cassette \#2, we first assembled and glued two intermediate stacks each composed of 4 small stacks of 25 mirrors. The rest of 240 mirrors was assembled in one run from small intermediate stacks. 

In Fig.~\ref{fig:CassetteScan}, positions 1-400 correspond to the first stack of 100 mirrors, positions 401-800 to the second stack of 100 mirrors, and positions 801-1600 to the rest composed of small stacks. In this two-step procedure, an error in the relative positioning of the second big stack would apply to all 100 mirrors in this stack. This fact may explain the asymmetric form of the slope distribution for Cassette \#2, or even the tendency to the splitting visible in Fig.~\ref{fig:CassetteScan}. 

In spite of this difference, the width of the slope distribution is in a good agreement with the tolerance window $\pm 1~mrad$ adopted during the assembling of both cassettes.

The angular distribution of individual mirror slopes in the cassette, $\sigma_\theta$, may be translated into the broadening of angular divergence of the incident neutron beam (given in second order):
\begin{equation}
{\rm FWHM}_{\rm eff} \approx {\rm FWHM}_{\rm in} \left( 1+\frac{1}{2}\left( \frac{2.35\sigma_\theta}{{\rm FWHM}_{\rm in}}\right) ^2 \right).
\label{eq:FWHA}
\end{equation}
Here, ${\rm FWHM}_{\rm in}$ is the incident beam angular divergence. For the PF1B guide, the incident beam divergence depends on the neutron wavelength, ${\rm FWHM}_{\rm in}\propto \lambda$, while $\sigma_\theta$ is, as purely geometrical effect, independent of $\lambda$. Therefore, we expect that the effective beam broadening from the polarizer is stronger for short wavelengths. 

The relevant wavelength band of the PF1B polarizer is $0.3-2.0~nm$. For neutrons with $\lambda =0.3~nm$, ${\rm FWHM}_{\rm PF1B}\approx 2 m \theta_{Ni} \lambda \approx 20~mrad$ (with $m = 2$ of the H113 SM coating and the critical angle of Ni $\theta_{Ni} =17.3~mrad/nm$ ~\cite{Hay1978jpe}). The effect of additional broadening due to the dispersion of mirror slopes, $\sigma_\theta\approx 1~mrad$, is expected to be about $0.7\%$ for Cassette \#2 and smaller for Cassette \#1, and even smaller for neutrons with longer wavelengths.

Comparing our results (Eq.~(\ref{eq:sigmas})) to the result $\sigma_{\rm coll}\approx 11~mrad$ measured for a stack of 200 Si plates of $0.2~mm$ thickness each, assembled without any control of angular orientation of individual plates ~\cite{Pet2019rsi}, demonstrates the importance and effectiveness of controlling the alignment of individual plates. Here, the plates are coated on both sides with Ti/Gd multi-layers and serve as collimator in a solid-state Fermi chopper. Since ${\rm FWHM}_{\rm coll}$ is comparable to the divergence after a neutron guide, such a collimator causes both, an imperfect collimation because of the transmission of neutrons outside the design divergence of the collimator, and a significant reduction of the on-axis beam intensity because of the absorption of low-divergence neutrons in the coating of misaligned plates.

\subsection{\label{sec:Final assembling}Final assembling of the polarizer}

\begin{figure}
\centering
\includegraphics[width=\columnwidth]{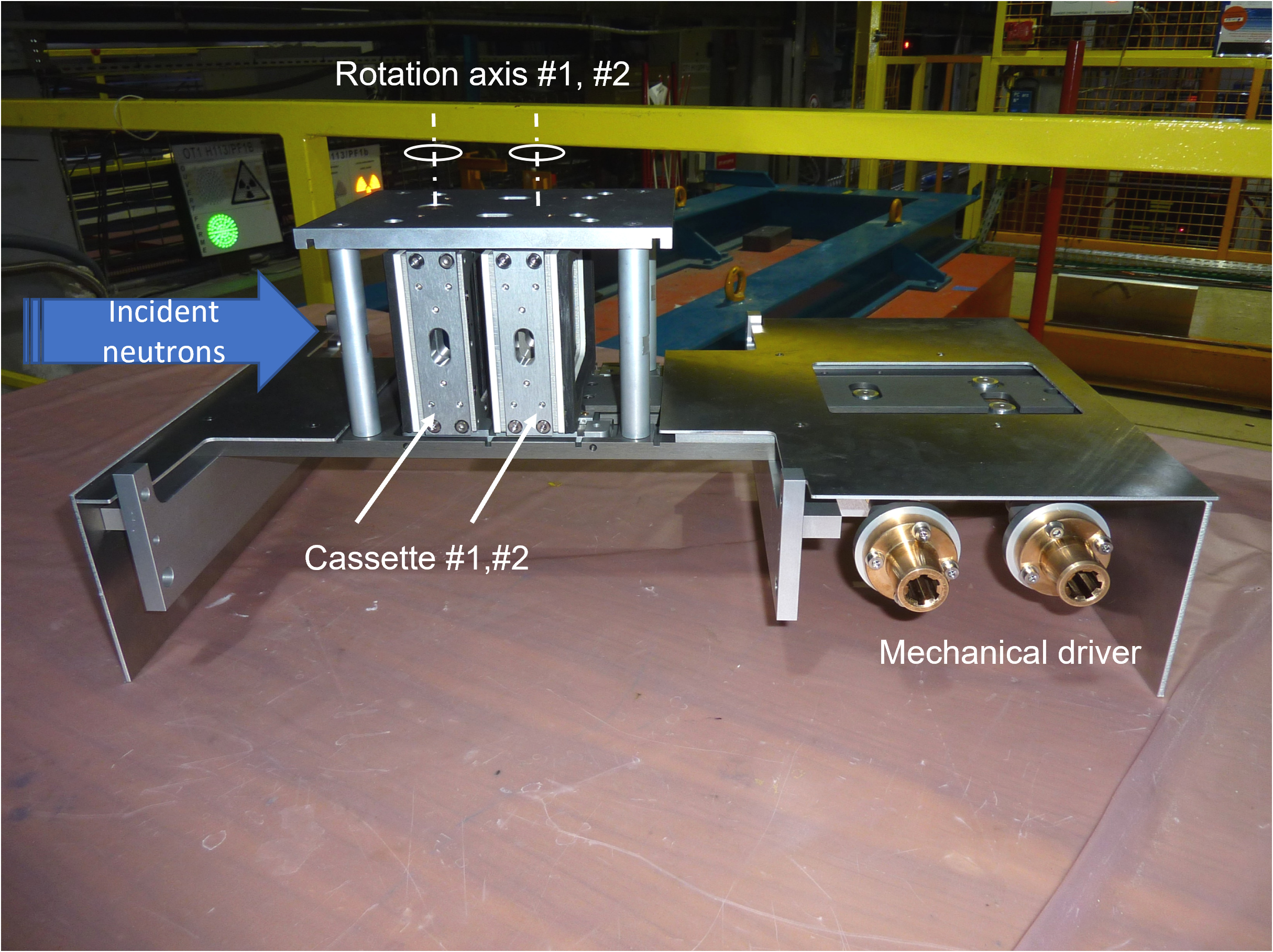}
\caption{\label{fig:FinalAssembly} The two polarizing cassettes mounted on the motorized mechanical driver. The motors and encoders are not visible, as they are placed outside the lead shielding where the polarizer is installed. From the motor axes, two ribbed shafts go through holes in the shielding and connect to the mechanics via the two brass hubs visible on the photo.}
\end{figure}

The two stacks of polarizing mirrors, Cassettes \#1 and \#2, are installed into a mechanical driver one after the other, see Fig.~\ref{fig:FinalAssembly}. The driver allows remote control of each stack's slope angle with respect to the direction of the neutron beam. The driver rotates both cassettes together around axis \#1 (centered at Cassette \#1) and separately Cassette \#2 around axis \#2 (centered at Cassette \#2). The angle between the two cassettes corresponds to the tilt angle $\gamma$, see Fig.~\ref{fig:Vbender}. Here and below we denote this angle $\gamma$, not $\gamma_V$, as we discuss only the V-bender below.  

\begin{figure*}
\centering
\includegraphics[width=\textwidth]{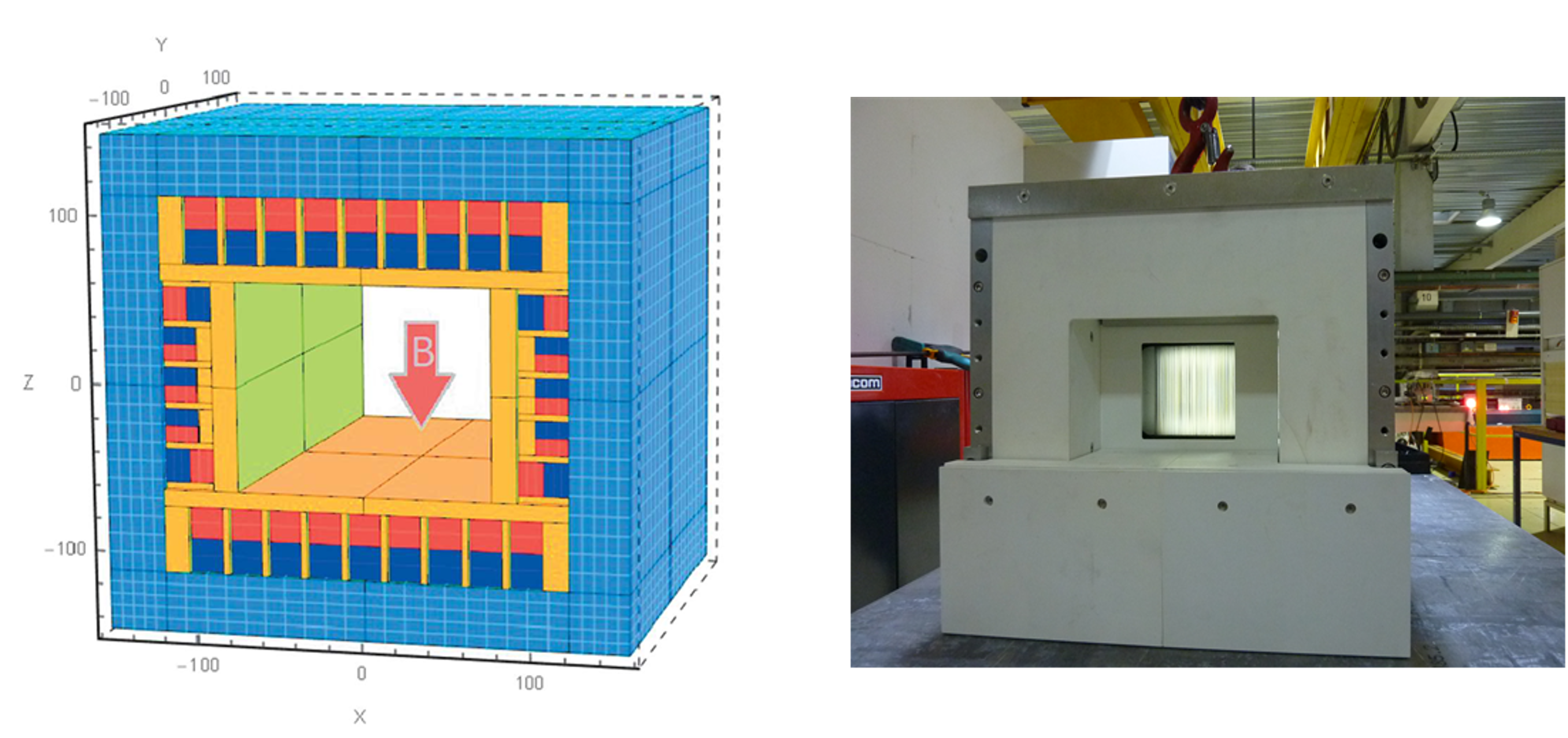}
\caption{\label{fig:Magnet} Left: Sketch of the magnetic housing for the SM polarizer of cold neutrons. Red color indicates the South poles and dense blue color the North poles of the NdFeB magnets~ \cite{Pet2019rsi}. Right: The fully assembled new PF1B polarizer composed of the magnet, the motorized mechanical insert with two polarizing cassettes, and the neutron shield built from boron nitride ceramics. The stacks were illuminated from behind and the light going through the sapphire plates can be seen.}
\end{figure*}

Then this assembly was inserted into the opening of the magnetic housing shown in Fig.~\ref{fig:Magnet}, Left. It is made from permanent magnets (for the details, see Ref. ~\cite{Pet2019rsi}) and provides a very homogeneous vertical field magnetizing the polarizer, with the transversal component $B_x/B_z<0.005$ in most of the volume occupied by the polarizing mirrors and $B_z\approx 0.38~T$. 

Finally, neutron shielding composed from boron nitride ceramics was mounted on both sides of the cassettes in order to absorb neutrons beyond the aperture of the polarizing mirrors, see Fig.~\ref{fig:Magnet}, right.

\section{\label{sec:Characterization}Characterization of the new polarizer}

\subsection{\label{sec:Setup}Experimental setup}

\begin{figure*}
\centering
\includegraphics[width=\textwidth]{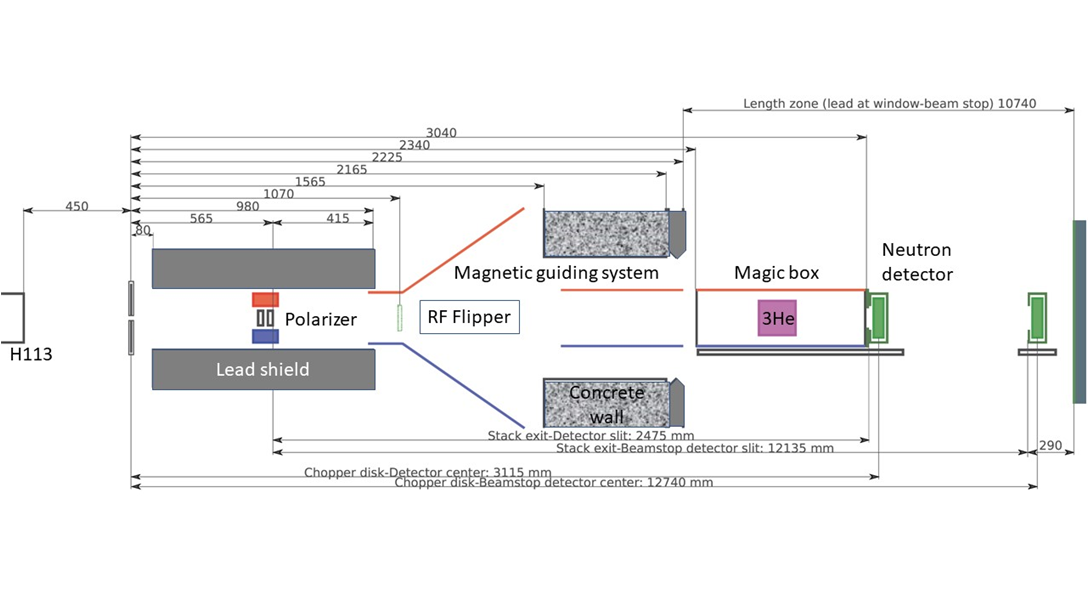}
\caption{\label{fig:Setup} Scheme of the experimental setup for the characterization of the new polarizer at PF1B.}
\end{figure*}

The fully assembled polarizer with its magnetic housing was installed into the lead shielding in the PF1B casemate, downstream the H113 neutron guide, see Fig.~\ref{fig:Setup}. 

An aperture of $70\times 70~mm^2$ was installed at the exit of the H113 neutron guide (where only $60~mm$ width are illuminated with neutrons). In order to measure time-of-flight (ToF) spectra, a neutron beam chopper with a horizontal slit of $60~mm$ width and $5~mm$ height was installed at a distance of $565~mm$ in front of the polarizer.

An adiabatic fast passage (AFP) neutron spin flipper with a flipping efficiency of $f>0.999$ ~\cite{Kre2005nima} was inserted in the space between the lead housing and the casemate wall. The stray field from the polarizer magnetic housing, the static magnetic field from the AFP flipper, a vertical magnetic field installed in the casemate window, and the magnetic field of the ``Magic box'' ~\cite{Pet2006nima} constitute the guiding field needed to transport adiabatically the neutron polarization. The ``Magic box'' also serves to conserve or flip the $^3$He polarization of the spin filter cell used for polarization analysis.

Both the ``Magic box'' and the neutron detector at its exit were installed on a motorized table in order to allow a horizontal scan of the beam. Neutron ToF measurements without spin filter cell were performed with a low-efficiency ($\sim 5\cdot 10^{-5}$) $^3$He detector. At low efficiency, the number of neutrons crossing the detector is weighted with the $^3$He cross section which grows linearly with neutron wavelength $\lambda$. Measurements with spin filter cell installed were performed with a detector with $\sim 3$\%\ efficiency. Where necessary, the height of the aperture in front of the detector was adapted to keep the dead time correction small.

In this experiment, we did not flip the neutron beam polarization and used the AFP flipper only to provide a static guiding field. Instead, we performed the polarization analysis by flipping the polarization of the  gas in the $^3$He spin filter cell~\cite{Bab2007pb} (polarization losses per flip $\delta <10^{-5}$). 

To align the polarizer relative to the beam direction, we first set both cassettes to be parallel to each other. Then we installed a $5~mm$ wide diaphragm behind the chopper thus reducing the beam cross section to $5\times 5~mm^2$. A large area neutron monitor was mounted just behind the polarizer exit to capture the full angular divergence of the incoming beam. 

In this configuration, we rotate both cassettes around axis \#1 using the motorized driver \#1 (from $-1.5^\circ$ to $+1.5^\circ$, in steps of $0.1^\circ$). The angular position with the maximum count rate corresponds to both cassettes being parallel to the incoming neutron beam. Then, we scan the angular position of Cassette \#2 keeping unchanged the position of Cassette \#1, in order to correct for a potential initial misalignment between the cassettes. Again, the maximum count rate corresponds to the second cassette being parallel to the first one and both parallel to the incident beam. In the last step, Cassette \#2 was tilted by the angle $\gamma\approx 5d/L\approx 18~mrad$ to prevent ``direct view'', see Fig.~\ref{fig:Scheme}. 

\subsection{\label{sec:AngularDistribution}Angular distribution}

Since the polarizer reflection plane is horizontal, the polarizer preserves the vertical angular distribution in the incident neutron beam and modifies only the horizontal one. The latter was measured by means of installing a small pinhole, $5\times 5~mm^2$, at the distance of $50~mm$ behind the chopper, and measuring the count rate in different detector positions across the beam far away from the pinhole using a small-aperture detector ($5~mm$ wide). In this measurement, the neutron detector was installed behind the ``Magic box'' at the distance of $3115~mm$ from the chopper.

\begin{figure}
\centering
\includegraphics[width=\columnwidth]{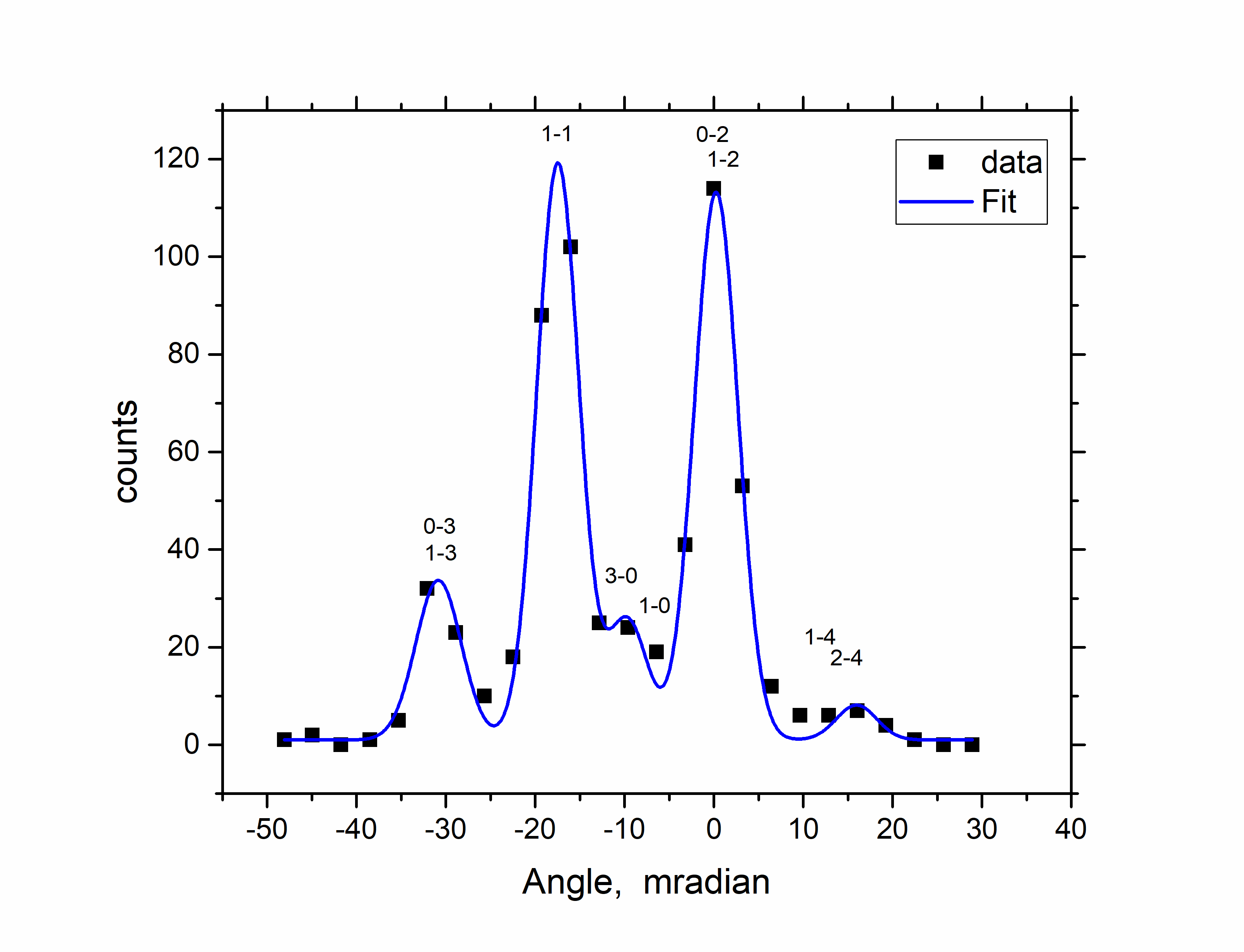}
\caption{\label{fig:ThreeHills}Angular distribution of neutrons in the horizontal plane behind the polarizer as measured with small apertures. The labels in the format ``a-b'' denote the number of reflections in Cassettes \#1 and \#2, respectively. Negative angles correspond to a detector displacement in the direction of bending with respect to the direct beam.}
\end{figure}

Fig.~\ref{fig:ThreeHills} shows the result of the angular scan for the polarizer tilt angle $\gamma\approx 5d/L$. Contrary to the commonly used C-bender, which shows a continuous angular distribution of reflected neutrons, the V-bender shows a discrete angular distribution for sufficiently large tilt angles. Each peak contains contributions from classes of neutron trajectories with a certain number of reflections in the first and in the second cassette. 

The labels in Fig.~\ref{fig:ThreeHills} denote the corresponding reflection numbers. For example, the label \mbox{1-1} corresponds to Garland trajectories with a single reflection in Cassette \#1 and a single reflection in Cassette \#2. The label \mbox{0-2} represents neutrons with Zig-Zag trajectories in Cassette \#2. Note that the position of the peak representing Zig-Zag \mbox{0-2} trajectories is independent of the tilt angle $\gamma$ and is close to the direction of the incident beam, while the positions of the other peaks strongly depend on $\gamma$. The solid line shows the result of multi-peak fitting with a Gaussian form. Trajectories with multiple reflections are suppressed by reflection losses, $R<100\%$, beyond the total reflection regime. 

The angular distribution measured with a point-like aperture on the source (in front of the polarizer as in the experiment or equivalently at the polarizer exit) has to be distinguished from the flux distribution $\Phi(x)$ measured for a large source:
\begin{equation}
\Phi(x)=\int\int \frac{\lambda}{\lambda_{th}}B{\rm d}\Omega {\rm d} \lambda,
\label{eq:Phi}
\end{equation}
where $B$ is the brightness of the neutron source ~\cite{Abe2006nima} and $\lambda_{\rm th}=0.18~nm$ is, by convention, the wavelength at the most probable velocity $v_0=2200~m/s$ in a thermal Maxwellian spectrum at the neutron temperature of $300~K$. Only in the ``near zone'' where the position splitting due to different angles in the beam is much smaller than the source size, a position scan provides a true flux density distribution. In contrast, a position scan in the ``far zone'' reproduces an angular distribution similar to Fig.~\ref{fig:ThreeHills}. This fact allows to profit from the full intensity of the beam in the ``near zone'' and to split the beam in well-collimated beams ($\sigma_\theta\approx 2.5~mrad$) in the ``far zone'' without any additional collimator. 

\subsection{\label{sec:Transmission}Polarizer transmission}

We used two independent methods to measure the integral transmission of the polarizer (integrated over all neutron wavelengths and angles). 

In the ToF method, we keep the chopper with its horizontal slit of $60\times 5~mm^2$ in place and compare intensities measured with the polarizer in and out of the beam (by translating the lead-shielding table holding the polarizer). The chopper slit is smaller than the polarizer of $80\times 80~mm^2$. Therefore, most of the neutrons passing through the chopper also hit the polarizer entrance (no angular collimation other than that from the width of $60~mm$ of the H113 guide was imposed in the horizontal plane). The $^3$He monitor detector (with an aperture of $20~mm$ width and $30~mm$ height) was installed behind the  ``Magic box'', see Fig.~\ref{fig:Setup}.

\begin{figure}
\centering
\includegraphics[width=\columnwidth]{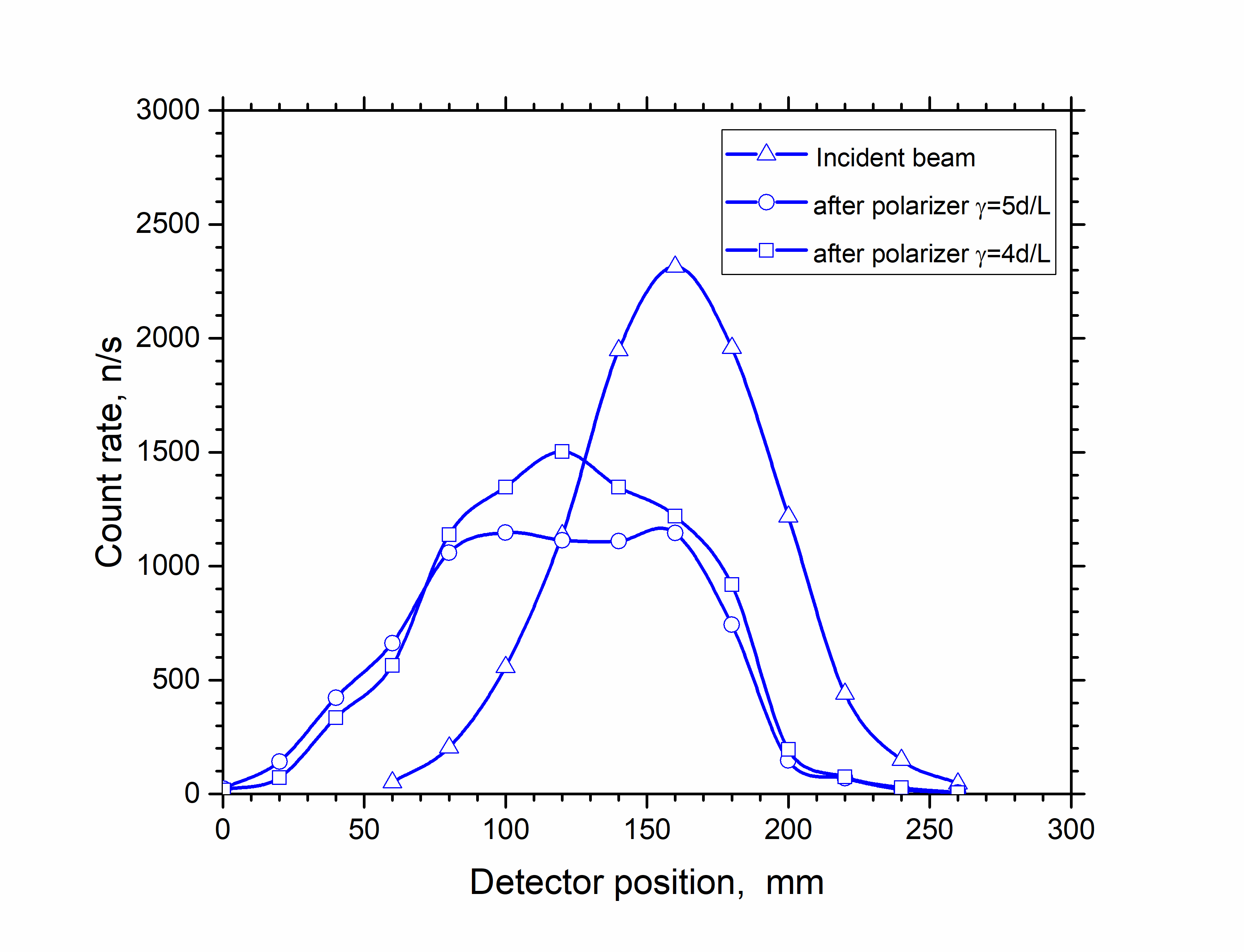}
\caption{\label{fig:HorizontalProfiles}Horizontal profiles of neutron capture flux measured at the distance of $3115~mm$ downstream the chopper: without polarizer (triangles) and downstream the polarizer for two different tilt angles, $\gamma\approx 4.5d/L$ (rectangles) and $\gamma\approx 5d/L$ (circles). Data with the polarizer in place have been scaled by a factor of 5.}
\end{figure}

We performed a horizontal scan of the beam by moving the detector with a step size of $20~mm$, corresponding to the width of the aperture, thus precisely mapping the horizontal axis, and measured ToF spectra at each detector position. Summing-up all channels in the ToF spectra we obtain the neutron count rate profiles shown in Fig.~\ref{fig:HorizontalProfiles}. Note that the data with polarizer in place were multiplied by a factor of 5. Integrating these data over all positions of the detector we found the polarizer integral transmission for the ``good'' spin component (i.e.\ relative to $1/2$ of the intensity of the unpolarized beam):
\begin{equation}
T_1=0.35~\text{for}~\gamma=4.5d/L~\text{, and}~
T_2=0.31~\text{for}~\gamma=5d/L,
\label{eq:Trans}
\end{equation}
where $\gamma$ is the polarizer tilt angle defined in Fig.~\ref{fig:Vbender}.

\begin{figure}
\centering
\includegraphics[width=\columnwidth]{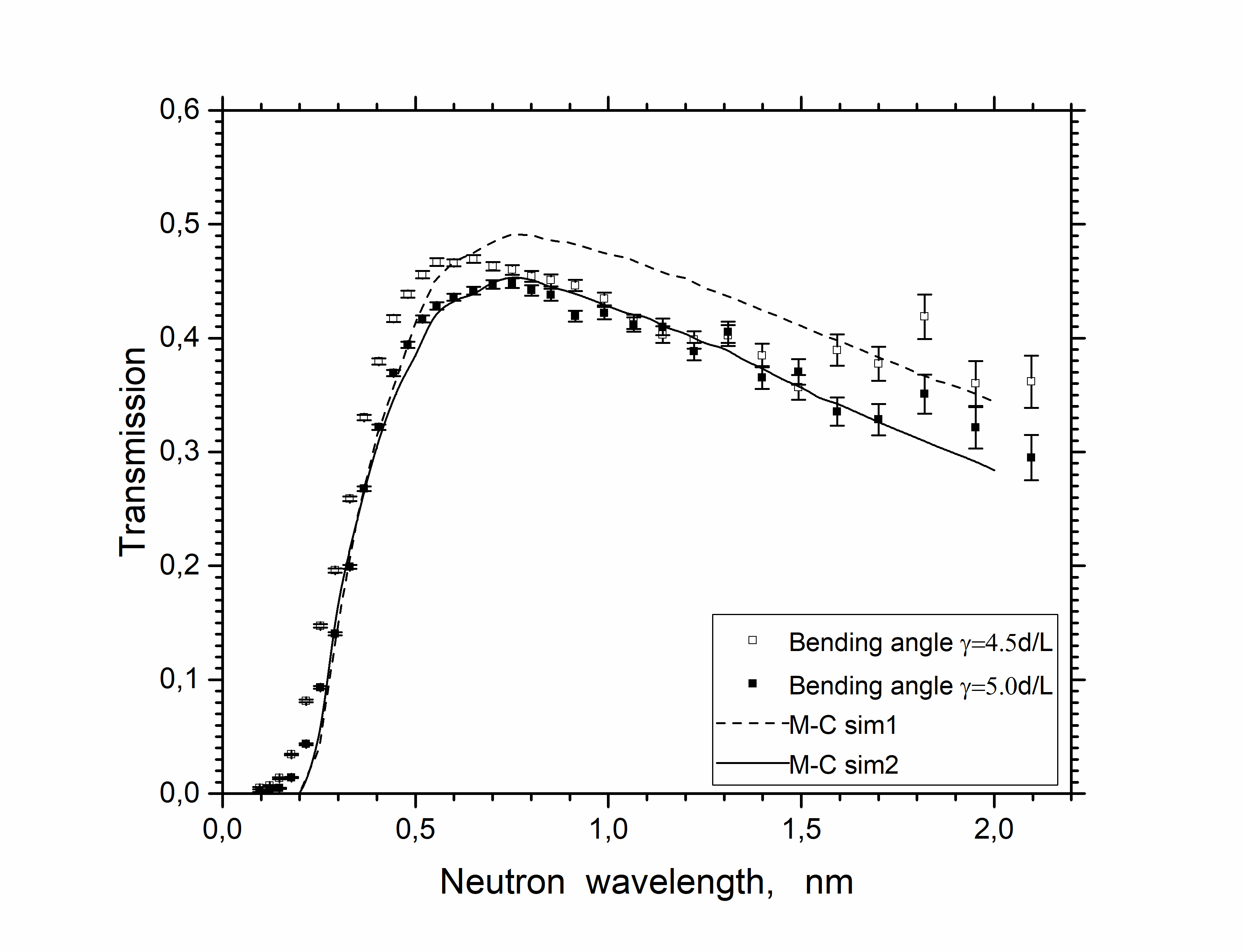}
\caption{\label{fig:Comparison} Comparison of the polarizer transmission for the ``good'' spin component measured at PF1B (points with error bars) and that simulated using a MC code (curves). The dashed line is obtained assuming a perfect alignment of the polarizer and reflectivity curve \#1 of Fig.~\ref{fig:Reflectivities}. The solid line corresponds to a small offset in $\gamma$, $\gamma=4.85d/L$, and reflectivity curve \#2 of Fig.~\ref{fig:Reflectivities}.}
\end{figure}

Using the same data without integration over all neutron wavelengths, we calculate the wavelength spectra of transmitted neutrons for the two polarizer tilt angles, see the experimental points with error bars in Fig.~\ref{fig:Comparison}. 

\begin{figure}
\centering
\includegraphics[width=\columnwidth]{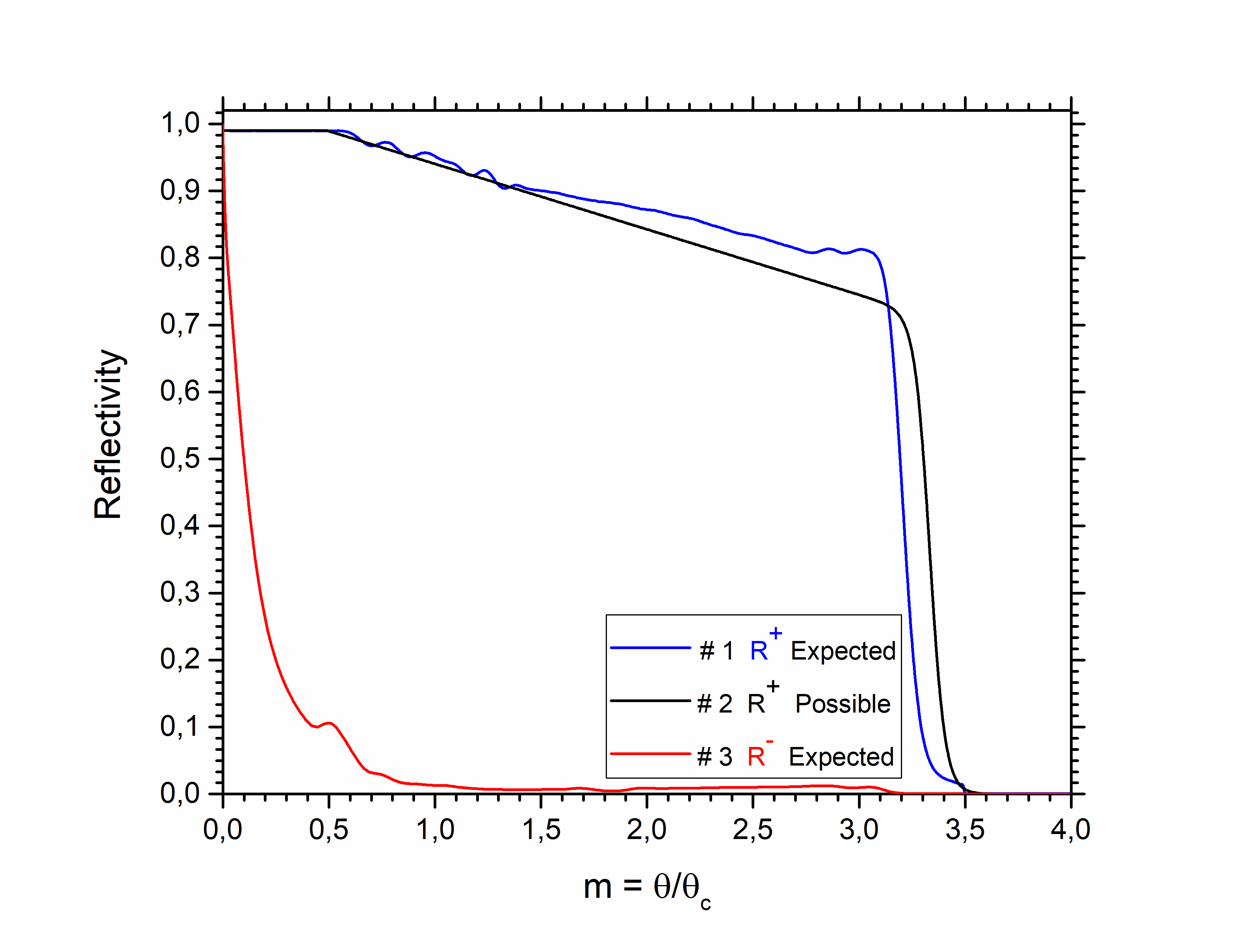}
\caption{\label{fig:Reflectivities} Reflectivity curves of the SM coating for the two neutron spin components, ($R^+$, $R^-$), propagating through a Sapphire substrate. The blue curve \#1 is from simulations performed in ~\cite{Pet2016nima,Pet2019rsi}. The black curve \#2 represents a parametrization ~\cite{Cla1997pb} with adjusted parameters to better match the experimental data in Fig.~\ref{fig:Performance}. See text for details.} 
\end{figure}

\begin{figure}
\centering
\includegraphics[width=\columnwidth]{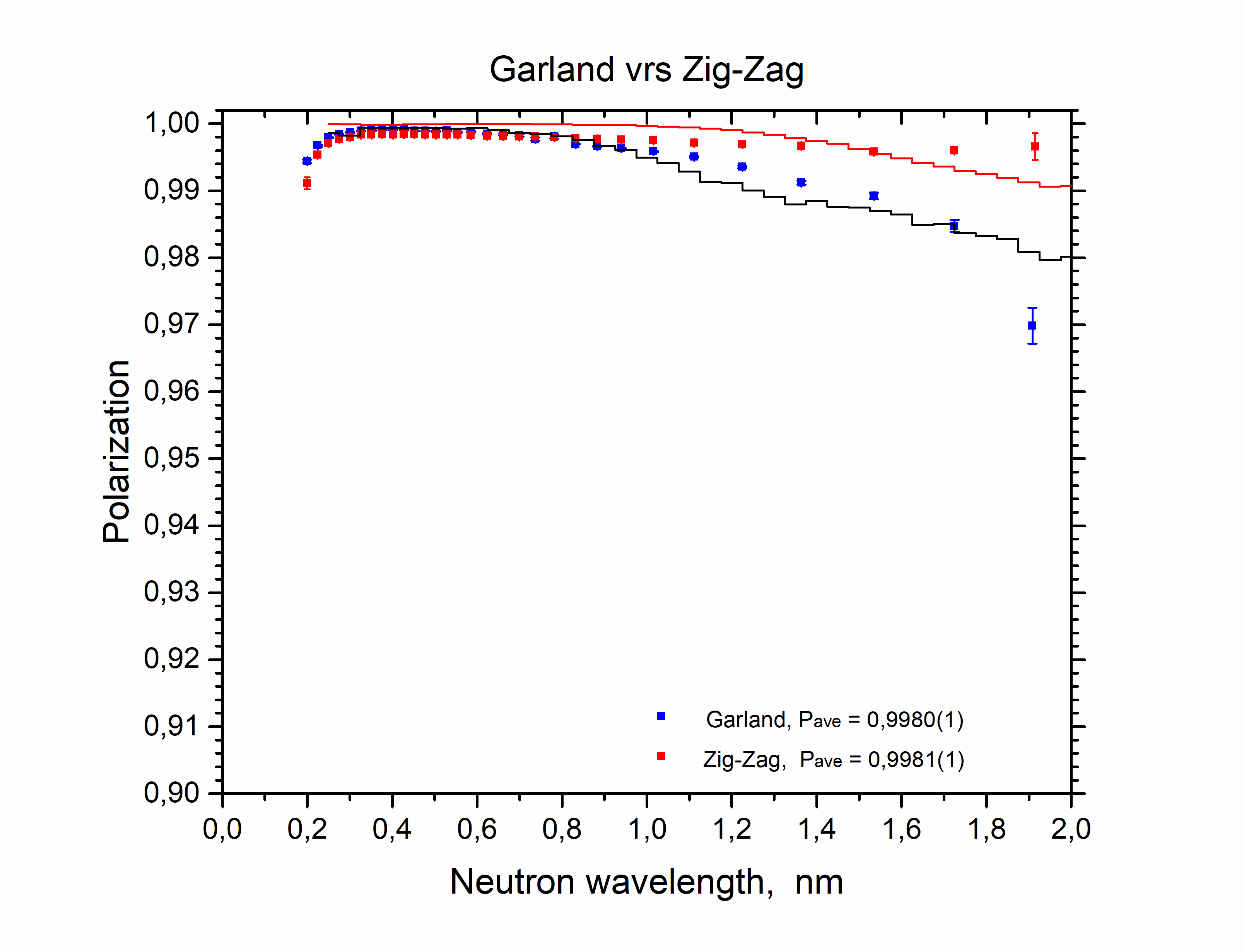}
\caption{\label{fig:Performance} Comparison of the polarizer performance for selected peaks in the output angular distribution measured at PF1B. Red color corresponds to neutrons with Zig-Zag trajectories, \mbox{0-2}, while blue color corresponds to Garland ones, \mbox{1-1}. Solid lines show a prediction of MC simulations. The decrease of measured polarization for short wavelengths, $\lambda <0.3~nm$, is due to the                   decrease of the $^3$He analyser opacity (which is shifted to longer wavelengths by the finite ToF resolution).} 
\end{figure}

First of all, we underline the excellent agreement between the experimental results and the results expected from simulation in the vicinity of the short wavelength cut-off. However, for longer wavelengths, the experimental transmission is systematically lower than the MC prediction; the relative difference is $\sim 10\%$. We do not know exactly the reason for this disagreement. Most probably, it is due to an uncertainty in our knowledge of the SM reflectivity curves for all 900 double-side coated plates as well as due to a dispersion in the mirrors' $m$-values, and due to an uncertainty in the angular alignment of the polarizer cassettes (the zero position found by searching for the maximum in the transmitted intensity has a finite precision). For example, a simulation performed with the slightly modified reflectivity curve shown by line \#2 in Fig.~\ref{fig:Reflectivities} and with a small offset in the tilt angle, $\sim 0.5~mrad$, gives a much better agreement, see the black solid line in Fig.~\ref{fig:Comparison}.

We also confirm that contrary to solid C-benders and S-benders, our V-bender does not show any Bragg dips in the wavelength transmission spectra.

In the second method of evaluation of the polarizer transmission, we measured the neutron capture flux in front and behind the polarizer, without other collimation than the aperture at the exit of the H113 guide, by activating thin Gold foils~\cite{Als1967nima} mounted on the polarizer entrance and exit windows, respectively. The results for the 5 foil positions arranged in a cross are given in Table~\ref{tab:Fluxes}.

\begin{table*}
\caption{\label{tab:Fluxes} Neutron capture fluxes in front and behind the polarizer, in units of $10^9 n/cm^2/s$. We performed all flux measurements at the reactor power of $56~MW$. The Mean values have been scaled to the nominal power of $58.3~MW$. The transmission is calculated for the ``good'' spin component ($1/2$ of the capture flux in front of the polarizer). The flux measured at the end of a collimation system frequently used for neutron decay experiments at PF1B is also given. This collimation system consists of a series of apertures of $6\times 6~cm^2$ installed in a vacuum tube, the first one just after the lead shield at $1.47~m$ and the last one at $4.87~m$ behind the guide exit, compare Fig.~\ref{fig:Scheme}.}
\begin{ruledtabular}
\begin{tabular}{cccccccc}
  Position:& Top & Centre & Bottom & Left & Right & Mean & Position, distance $x$ from guide exit\\ 
Flux:& 20.1 & 20.4 & 21.5 & 19.1 & 20.1 & 21.1 & Polarizer entrance, $x=0.87~m$ \\
Flux:& 3.65 & 3.53 & 3.63 & 3.21 & 3.12 & 3.57 & Polarizer exit, $x=1.07~m$ \\
Transmission [\%]: & 36.4 & 34.6 & 33.8 & 33.6 & 31.0 & 33.8 & \\\hline\hline 
Flux:& 0.59 & 0.64 & 0.61 & 0.63 & 0.66 & 0.65 & End of collimation, $x=4.87~m$ \\
\end{tabular}
\end{ruledtabular}
\end{table*}

The data in Table~\ref{tab:Fluxes} show good transmission homogeneity both in vertical and horizontal direction as well as a reasonable agreement with the results of the ToF method. The small difference between the results of the two methods may be explained by the difference in the angular acceptance: the result Eq.~(\ref{eq:Trans}) was obtained using nearly the total horizontal angular divergence of the beam, which is not fully correct for the results of the gold foil activation shown in Table~\ref{tab:Fluxes}. From the measurements of the integral transmission of the new polarizer as well as from measurements of the transmission as a function of neutron wavelength, we underline the good agreement between MC simulations and experimental data, which is not often the case for such kind of polarizers~\cite{Sha2014nima}.

\subsection{\label{sec:Power}Polarization performance}

The polarization of the neutron beam downstream the polarizer was measured using the setup shown in Fig.~\ref{fig:Setup}. The only difference compared with the transmission experiment is the presence of a cell with Si windows filled with polarized $^3$He gas (length of the gas column: $15~cm$), mounted inside the ``Magic box'' in order to preserve the polarization of the $^3$He. 

The polarized $^3$He was produced by Metastable Optical Pumping (MEOP) using the ILL filling station TYREX~\cite{And2005pb,Pet2006pb}. To minimize eventual systematic uncertainties in the measurements of neutron polarization we used the method of opaque cells~\cite{Zim1999plb}. The $^3$He polarization measured optically~\cite{Big1992jp} on TYREX was $0.75$. 

The polarization of the neutron beam behind the polarizer was analyzed by means of RF flipping the polarization of the $^3$He gas in the analyzer cell~\cite{Bab2007pb}. The loss in $^3$He polarization per single flip was $<10^{-5}$. 

To cover the neutron wavelength range of interest, $0.3-2.0~nm$, we used the following set of $^3$He pressures in the cell: $0.51,~0.82,~2.2~bar$. This set of ``opaque'' $^3$He cells provides $>0.999$ analyzing power for the neutron wavelengths of $>1.0,~0.6,~0.22~nm$, respectively. After filling the cell with polarized $^3$He, it is inserted in a compact magnetic transport system~\cite{Hay1978jpe} and transported to PF1B where it is installed in the ``Magic box''. The spin-relaxation time constant for $^3$He gas in the cell was longer than $200~h$~\cite{Pet2006pb}. Both applied techniques (opaque polarized $^3$He analyzer and in-situ RF flipping of the $^3$He spin state) assure $>0.999$ analyzing power for neutron spin analysis, without any corrections. 

\subsubsection{\label{sec:FarZone}Far zone}

As mentioned in Section~\ref{sec:AngularDistribution}, the neutron intensity distribution across the beam in the far zone resolves the angular distribution present after the polarizer. Therefore, we decided to measure the polarization for the two most intense peaks in this angular distribution. These peaks correspond to Garland and Zig-Zag trajectories in the polarizer, marked as \mbox{1-1} and \mbox{0-2} in Fig.~\ref{fig:ThreeHills}, respectively. With a small aperture, $5\times 5~mm^2$, installed just behind the chopper, we measured the neutron angular distribution by means of a detector position scan across the beam at the distance of $3.15~m$ from the chopper. The full range of the scan was $200~mm$ with a step size of $20~mm$; this range corresponds to the beam divergence of $3.7^\circ$. At each point of the scan, we measured ToF spectra for the ``white'' (high transmission) spin state of the $^3$He analyzer. With a detector horizontal aperture of $20~mm$, we were able to fully cover the selected peaks. Finally, we centered the detector to either of the two most intense beams and measured ToF spectra for the two spin states of the analyzer. The measurements were performed using the loop \{``white'', ``black'', ``black'', ``white''\} to minimize possible systematic effects associated with the slow decay of the $^3$He polarization. To partly compensate for the very different count rates for the two spin states of the analyzer (the raw flipping ratio is $\sim 10^3$) we used very different expositions for each analyzer state: \{$30~s$, $1800~s$, $1800~s$, $30~s$\}. Sufficient statistics was accumulated by repeating this sequence for $10-20~h$. In order to preserve a high efficiency of the $^3$He analyzer cell, we replaced the cell with a freshly filled one every $24-48~h$. Note that the neutron transmission of the spin filter cell may change significantly on this time scale, whereas the analyzing power remains stable in the region of interest, where it is in saturation ($A\rightarrow 1$). 

The measured polarization (rectangular points) for the Zig-Zag \mbox{0-2} (red points) and Garland \mbox{1-1} (blue points) trajectories are shown in Fig.~\ref{fig:Performance}. First of all, we note the excellent polarization for both selected peaks. For the neutron wavelength band of $0.3-0.6~nm$, corresponding to the intensity maximum of the polarized beam, the measured polarization reaches the value of $P_n\approx 0.999$. The polarization values averaged over all the transmitted spectra are: $P_n\approx 0.9981(1)$ for the \mbox{0-2} Zig-Zag trajectories and $P_n\approx 0.9980(1)$ for the \mbox{1-1} Garland trajectories, where only the statistical uncertainties are given.

As expected, the mean reflection angle for Zig-Zag trajectories is higher and, therefore, the cut-off wavelength is at a longer wavelength than for Garland trajectories. This feature of Zig-Zag trajectories also explains the practically constant polarization (red points) for the full wavelength band $\lambda\in\{0.3~nm-1.9~nm\}$.

\subsubsection{\label{sec:NearZone}Near zone}

\begin{figure}
\centering
\includegraphics[width=\columnwidth]{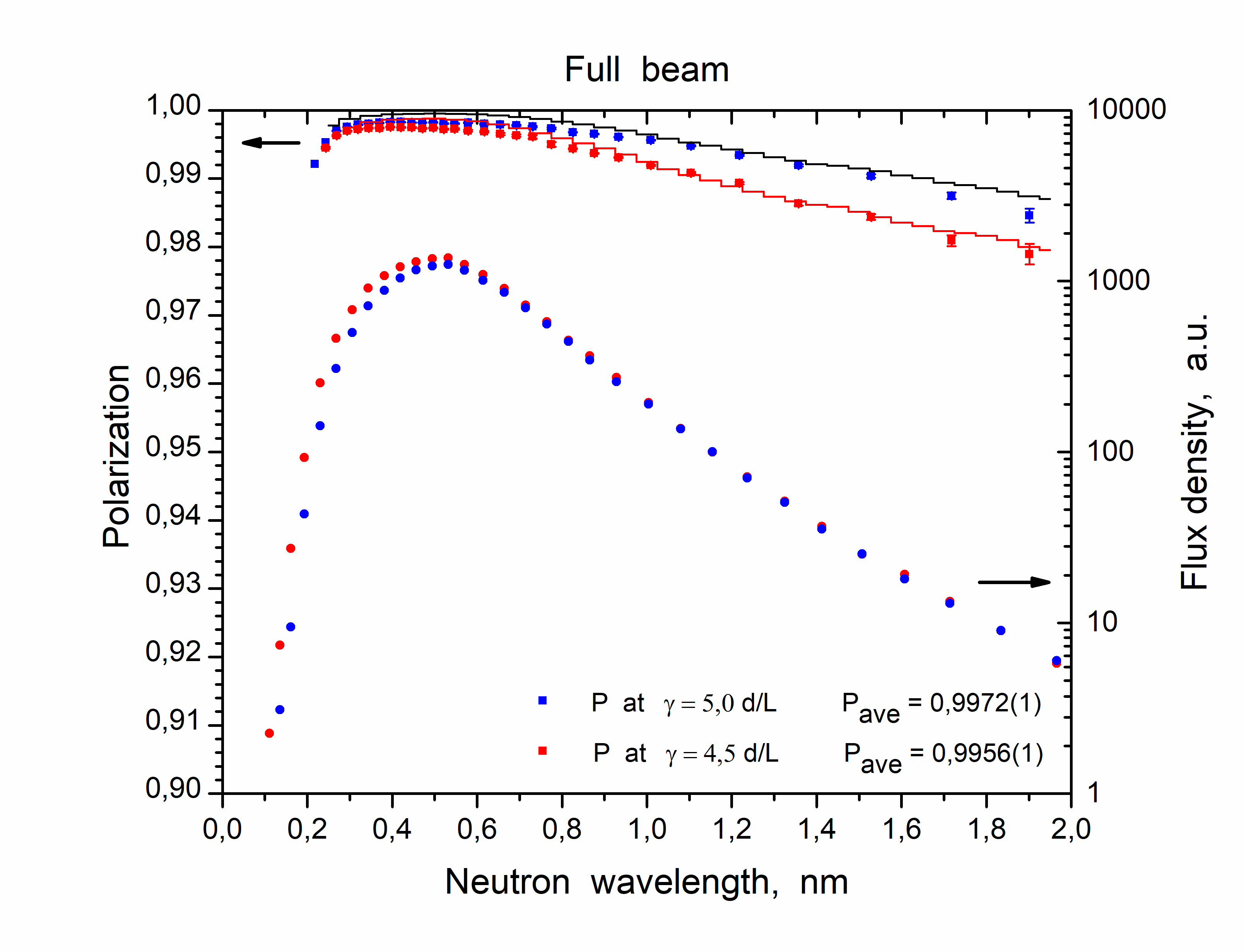}
\caption{\label{fig:ComparisonBis} Comparison of the polarizer performance for two different tilt angles $\gamma$ (Red: $\gamma=4.5~d/L$, Blue: $\gamma=5~d/L$). The rectangles with error bars represent the wavelength spectra of neutron polarization (left axis) and the circles the wavelength spectra of capture flux intensity (right axis). The decrease of the measured polarization for short wavelengths, $\lambda<0.3~nm$, is due to the decrease of the $^3$He analyzer opacity (which is shifted to longer wavelengths by the finite ToF resolution).} 
\end{figure}

In the near zone, the value of interest is the polarization averaged over the full beam (fully illuminated polarizer and all angles of transmitted neutrons). Since the polarizer does not modify neutron trajectories in the vertical plane, we don't expect any variation of the neutron polarization in the vertical direction. Therefore, we used the full horizontal aperture of the chopper ($60\times5~mm^2$) and measured ToF spectra for the two spin states across the full transmitted beam in the horizontal plane. In order to precisely map the beam over its width of $200~mm$, it was scanned with a step size corresponding to the width of the detector aperture, $20~mm$. The data obtained after integration over all positions of the detector at the distance of $3115~mm$ from the chopper are shown in Fig.~\ref{fig:ComparisonBis}.

Again, we note the outstanding polarizing power of the new PF1B polarizer even after averaging over the full output phase-space (angle, position, wavelength). It was measured as high as $P_n=0.9960(1)$ for $\gamma=4.5~d/L$ ($16.2~mrad$) and $P_n=0.9974(1)$ for $\gamma=5~d/L$ ($18~mrad$) (where the measured capture flux spectra are used for averaging over the wavelengths).

\section{\label{sec:Adaptability}Polarizer Adaptability}

As mentioned in Section ~\ref{sec:Final assembling}, the new PF1B polarizer is equipped with two motorized drivers, see Fig.~\ref{fig:FinalAssembly}. The V-shape design of the polarizer and the motorization provide the unique opportunity to control remotely the polarizer orientation relative to the incident neutron beam as well as the tilt angle $\gamma$ between the polarizing cassettes (analogous to the bending angle of a classical C-bender). 

The commonly accepted criterion for choosing the value of the tilt angle is the critical angle $\gamma_c$ which just prohibits the ``direct view''. In other words, it is the minimal angle which guarantees the absence of neutron trajectories without collisions with the polarizer reflecting mirrors. One may ask the question: is this angle optimal for all types of experiments with polarized neutrons?

Obviously, this is not the case. Indeed, for a long instrument downstream the polarizer, one is interested in the on-axis value $\partial_\Omega\Phi$ ($\Phi$ is the neutron flux density in the beam). An example is the PERKEO~II experiments~\cite{Kre2005plb,Mund:2012fq} which used a well-collimated beam with rather low divergence acceptance. In contrast, for a short instrument, which accepts a high beam divergence, the quantity of interest is the integral flux density $\int_A\Phi{\rm d}A$. These two situations correspond to Far and Near zone after the polarizer.

If considering systematic effects, a very wide class of experiments is statistically limited and systematic uncertainties associated with a spatial or angular non-uniformity of the polarization are not dominant~\cite{Ves2008prc,Gle2017plb,Goe2007plb,Gag2016prc}. For this class of experiments, the so-called Figure-of-Merit (FoM): $\Lambda=P^2T$ is the quantity of interest. The bending angle $\gamma_c$ defined above does not maximize $\Lambda$. On the other hand, experiments which are extremely sensitive to the polarization distribution over the beam cross section or over the emitting angle, profit from an ultra-high beam polarization which leaves no room for noticeable systematic uncertainties. 

The data shown in Fig.~\ref{fig:ComparisonBis} were measured for two different tilt angles $\gamma=4.5d/L$ and $\gamma=5d/L$, which are $12\%$ and $20\%$ higher than $\gamma_c=4d/L$ of the ``no direct view'' condition. 

The lower the angle $\gamma$ the lower is the beam polarization and the higher is the polarizer transmission or the beam flux density $\Phi$. This typical concurrence implies the existence of an optimal angle $\gamma$ which maximizes the FoM value. There is no reason to expect this optimal value $\Lambda$ to coincide with the value resulting from the ``no direct view'' condition. With the values $P_{\rm ave}$ and $T_{\rm ave}$ shown in Fig.~\ref{fig:ComparisonBis}, we arrive at the following value of the $\Lambda$ parameter: for $\gamma=4.5d/L$, $\Lambda=0.346$, and for $\gamma=5d/L$, $\Lambda=0.312$. This means that even a lower value of $\gamma$ is required to maximize the parameter $\Lambda$. Note that the tilt angle $\gamma=5d/L$ was chosen to provide ultra-high beam polarization, corresponding to what was achieved with the X-SM geometry of two of the previous PF1B benders, where it improves the FoM value substantially compared to $\Lambda_{\rm X-SM}=0.24$ of the X-SM geometry.

\begin{figure}
\centering
\includegraphics[width=\columnwidth]{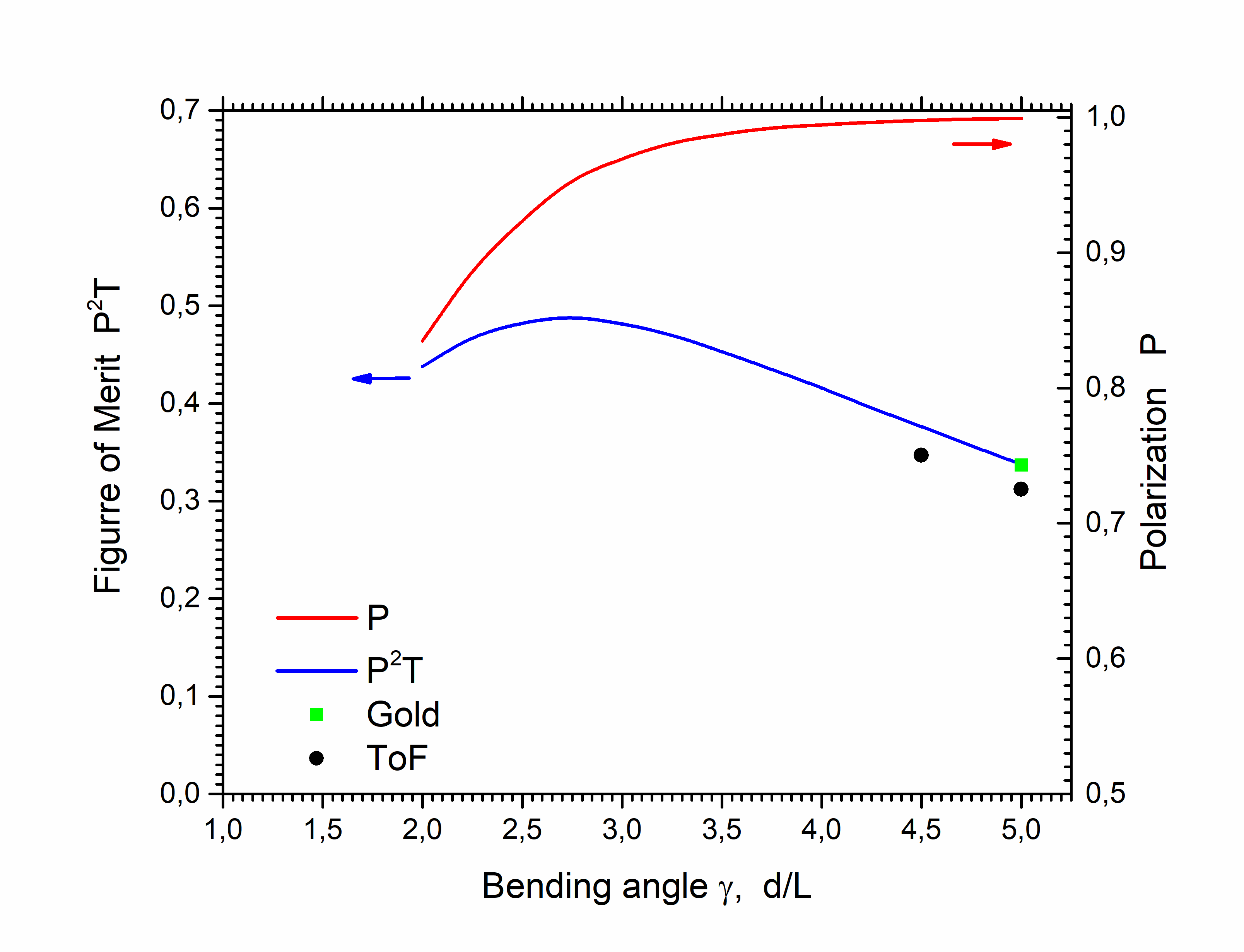}
\caption{\label{fig:Merit} Figure-of-Merit $\Lambda$ (blue, left axis) and polarization (red, right axis) as function of the polarizer tilt angle $\gamma$ in units $d/L$. The solid lines show results of MC simulations. The points mark our experimental results for $\Lambda$ (green: transmission from gold foil activation, black: transmission from ToF data). Note that the experimental points are for tilt angles well above the ``no direct view'' condition where the performance of the real device, in particular its polarization, is rather insensitive to small geometrical imperfections. For tilt angles in the transition region these imperfections have to be taken into account.}
\end{figure}

Since measurements of the polarizer transmission and especially the polarization are time consuming, we did not perform a full scan over the tilt angle. Instead, we tried to shed a light on this problem using results of MC ray tracing, see Fig.~\ref{fig:Merit}.   
One can see that for the V-bender geometry, the maximum of $\Lambda$ (blue solid line) is reached for tilt angles of about $\gamma=2.5-3.0d/L$ which is far below the value $\gamma=4.0d/L$ required to avoid the ``direct view''. Comparing the $\Lambda$ curve in Fig.~\ref{fig:Merit} with the value $\Lambda=0.477$ measured for the previous PF1B polarizer ~\cite{Sol2002ILL} we conclude that the new polarizer at a tilt angle $\gamma\approx 2.75d/L$ delivers the same Figure-of-Merit $\Lambda$ as the previous one. 

This fact opens the possibility to optimize the polarized beam for an extremely broad class of possible experiments using the same polarizing device: either for ultra-high polarization, or for the highest Figure-of-Merit.

More generally, these results show that polarizing devices similar to the one presented here could replace advantageously conventional benders in other types of neutron instruments, even for experiments that are typically flux-limited but do not require the highest level of polarization. These devices give the possibility to tune the polarizer performance for particular experiments depending on its critical requirement (polarization or transmission). More specifically, for neutron scattering experiments where only limited divergence can be used, one could select only the output peak \mbox{0-2} in Fig.~\ref{fig:ThreeHills} where the intensity corresponding to ``zig-zag'' neutron trajectories is concentrated, close to the incident beam direction, with  polarization $0.995-0.999$ over the whole wavelength range (Fig.~\ref{fig:Performance}). By contrast, a C-shaped bender with similar parameters would deviate the whole beam more strongly, giving a continuous outgoing angular distribution~\cite{Soy1995pb} whose total width would be comparable to the present V-bender case~\cite{Pet:un}. The polarizing device could be used either on continuous reactor-based sources or pulsed sources (e.g. spallation), as our ToF measurements show that it can be operated over a broad wavelength range. Another application of this kind of device would be to analyze precisely the beam polarization at a given instrument, without ambiguity and without sophisticated data treatment (assuming that the beam divergence is smaller than the acceptance of the analyzer). Due to the high analyzing efficiency, it could be used the same way as an opaque $^3$He spin filter cell ~\cite{Zim1999plb}, in case the latter is not available or applicable and that an accuracy of a few times $10^{-3}$ is sufficient. A device with smaller beam cross section, easier to install on most instruments, could be built and made available on-demand.
 
\section{\label{sec:Conclusion} Conclusion}

A new type of solid-state polarizer is built entirely in-house at the ILL for the PF1B cold neutron beam facility. The polarizer is installed in the PF1B casemate and tested in real conditions. 

For the tilt angle $\gamma=5 d/L$, in the near zone, the downstream polarization, averaged over the capture spectrum for the full wavelength band of $0.3-2.0~nm$, reaches a record measured value of $P_n=0.997(1)$, with a mean transmission for the ``good'' spin component $>0.31$. In the far zone, the polarization is $P_n>0.998$, and it is practically independent of the neutron wavelength within the wavelength band $\lambda=0.3-1.0~nm$, which contains $0.97$ of the total flux. For longer wavelengths, the polarization shows a very slow decrease towards $0.995$ at the wavelength $\lambda=2~nm$.

The polarizer allows remote control of its geometry (the tilt angle $\gamma$) which opens an unique option to deliver optimal conditions for an extremely broad class of possible experiments using the same polarizing device: either ultra-high polarization or optimal Figure-of-Merit, or the option to adjust the mean take-off angle.

To our knowledge, no device with similar performance endorsed by measurements was reported for a cold neutron polarizing device. The polarizer is based on a series of innovations in the design and fabrication in the following domains: choice of the substrate material, SM and anti-reflecting multilayer coatings, strength and homogeneity of the magnetizing field, and precision of the assembling process.

The polarizer is used for user experiments since the last reactor cycle in 2020.

\begin{acknowledgments}
These measurements were performed at the instruments PF1B, SuperADAM, D17, and T13 at the ILL. We express our gratitude to our technicians Didier Berruyer, Pascal Mouveau and Nicolas Thiery who produced many technical components for the experiments, to Nicolas Surget and Benjamin Sornin for the new control system of the SM coating machine, and to Guillaume Delphin, Amandine Vittoz, Florian Philit and Vincent Gaignon for the production of the SMs by magnetron sputtering. We thank Thomas Saerbeck for useful transmission and reflectivity measurements of substrate materials.
\end{acknowledgments}

\section*{Data Availability Statement}

Data openly available in a public repository that issues datasets with DOIs: doi:10.5291/ILL-DATA.TEST-3099, doi:10.5291/ILL-DATA.3-07-366, doi:10.5291/ILL-DATA.TEST-2541.

\section{References}
\bibliography{aipsamp}

\providecommand{\noopsort}[1]{}\providecommand{\singleletter}[1]{#1}%
\begin{thebibliography}{49}%
\makeatletter
\providecommand \@ifxundefined [1]{%
 \@ifx{#1\undefined}
}%
\providecommand \@ifnum [1]{%
 \ifnum #1\expandafter \@firstoftwo
 \else \expandafter \@secondoftwo
 \fi
}%
\providecommand \@ifx [1]{%
 \ifx #1\expandafter \@firstoftwo
 \else \expandafter \@secondoftwo
 \fi
}%
\providecommand \natexlab [1]{#1}%
\providecommand \enquote  [1]{``#1''}%
\providecommand \bibnamefont  [1]{#1}%
\providecommand \bibfnamefont [1]{#1}%
\providecommand \citenamefont [1]{#1}%
\providecommand \href@noop [0]{\@secondoftwo}%
\providecommand \href [0]{\begingroup \@sanitize@url \@href}%
\providecommand \@href[1]{\@@startlink{#1}\@@href}%
\providecommand \@@href[1]{\endgroup#1\@@endlink}%
\providecommand \@sanitize@url [0]{\catcode `\\12\catcode `\$12\catcode
  `\&12\catcode `\#12\catcode `\^12\catcode `\_12\catcode `\%12\relax}%
\providecommand \@@startlink[1]{}%
\providecommand \@@endlink[0]{}%
\providecommand \url  [0]{\begingroup\@sanitize@url \@url }%
\providecommand \@url [1]{\endgroup\@href {#1}{\urlprefix }}%
\providecommand \urlprefix  [0]{URL }%
\providecommand \Eprint [0]{\href }%
\providecommand \doibase [0]{http://dx.doi.org/}%
\providecommand \selectlanguage [0]{\@gobble}%
\providecommand \bibinfo  [0]{\@secondoftwo}%
\providecommand \bibfield  [0]{\@secondoftwo}%
\providecommand \translation [1]{[#1]}%
\providecommand \BibitemOpen [0]{}%
\providecommand \bibitemStop [0]{}%
\providecommand \bibitemNoStop [0]{.\EOS\space}%
\providecommand \EOS [0]{\spacefactor3000\relax}%
\providecommand \BibitemShut  [1]{\csname bibitem#1\endcsname}%
\let\auto@bib@innerbib\@empty
\bibitem [{\citenamefont {Abele}\ \emph {et~al.}(2006)\citenamefont {Abele},
  \citenamefont {Dubbers}, \citenamefont {Hase}, \citenamefont {Klein},
  \citenamefont {Knopfler}, \citenamefont {Kreuz}, \citenamefont {Lauer},
  \citenamefont {M{\"a}rkisch}, \citenamefont {Mund}, \citenamefont
  {Nesvizhevsky}, \citenamefont {Petukhov}, \citenamefont {Schmidt},
  \citenamefont {Schumann},\ and\ \citenamefont {Soldner}}]{Abe2006nima}%
  \BibitemOpen
  \bibfield  {author} {\bibinfo {author} {\bibfnamefont {H.}~\bibnamefont
  {Abele}}, \bibinfo {author} {\bibfnamefont {D.}~\bibnamefont {Dubbers}},
  \bibinfo {author} {\bibfnamefont {H.}~\bibnamefont {Hase}}, \bibinfo {author}
  {\bibfnamefont {M.}~\bibnamefont {Klein}}, \bibinfo {author} {\bibfnamefont
  {A.}~\bibnamefont {Knopfler}}, \bibinfo {author} {\bibfnamefont
  {M.}~\bibnamefont {Kreuz}}, \bibinfo {author} {\bibfnamefont
  {T.}~\bibnamefont {Lauer}}, \bibinfo {author} {\bibfnamefont
  {B.}~\bibnamefont {M{\"a}rkisch}}, \bibinfo {author} {\bibfnamefont
  {D.}~\bibnamefont {Mund}}, \bibinfo {author} {\bibfnamefont {V.~V.}\
  \bibnamefont {Nesvizhevsky}}, \bibinfo {author} {\bibfnamefont
  {A.}~\bibnamefont {Petukhov}}, \bibinfo {author} {\bibfnamefont
  {C.}~\bibnamefont {Schmidt}}, \bibinfo {author} {\bibfnamefont
  {M.}~\bibnamefont {Schumann}}, \ and\ \bibinfo {author} {\bibfnamefont
  {T.}~\bibnamefont {Soldner}},\ }\bibfield  {title} {\enquote {\bibinfo
  {title} {Characterization of a ballistic supermirror neutron guide},}\
  }\href@noop {} {\bibfield  {journal} {\bibinfo  {journal} {Nucl. Instrum.
  Meth. A}\ }\textbf {\bibinfo {volume} {562}},\ \bibinfo {pages} {40}
  (\bibinfo {year} {2006})}\BibitemShut {NoStop}%
\bibitem [{\citenamefont {Vesna}\ \emph {et~al.}(2008)\citenamefont {Vesna},
  \citenamefont {Gledenov}, \citenamefont {Nesvizhevsky}, \citenamefont
  {Petoukhov}, \citenamefont {Sedyshev}, \citenamefont {Soldner}, \citenamefont
  {Zimmer},\ and\ \citenamefont {Shul'gina}}]{Ves2008prc}%
  \BibitemOpen
  \bibfield  {author} {\bibinfo {author} {\bibfnamefont {V.~A.}\ \bibnamefont
  {Vesna}}, \bibinfo {author} {\bibfnamefont {Y.~M.}\ \bibnamefont {Gledenov}},
  \bibinfo {author} {\bibfnamefont {V.~V.}\ \bibnamefont {Nesvizhevsky}},
  \bibinfo {author} {\bibfnamefont {A.~K.}\ \bibnamefont {Petoukhov}}, \bibinfo
  {author} {\bibfnamefont {P.~V.}\ \bibnamefont {Sedyshev}}, \bibinfo {author}
  {\bibfnamefont {T.}~\bibnamefont {Soldner}}, \bibinfo {author} {\bibfnamefont
  {O.}~\bibnamefont {Zimmer}}, \ and\ \bibinfo {author} {\bibfnamefont {E.~V.}\
  \bibnamefont {Shul'gina}},\ }\bibfield  {title} {\enquote {\bibinfo {title}
  {Measurement of the parity violating triton emission asymmetry in the
  reaction {$^6Li(n,\alpha)^3H$} with polarized cold neutrons},}\ }\href@noop
  {} {\bibfield  {journal} {\bibinfo  {journal} {Phys. Rev. C}\ }\textbf
  {\bibinfo {volume} {77}},\ \bibinfo {pages} {035501} (\bibinfo {year}
  {2008})}\BibitemShut {NoStop}%
\bibitem [{\citenamefont {Gledenov}\ \emph {et~al.}(2017)\citenamefont
  {Gledenov}, \citenamefont {Nesvizhevsky}, \citenamefont {Sedyshev},
  \citenamefont {Shul'gina}, \citenamefont {Szalanski},\ and\ \citenamefont
  {Vesna}}]{Gle2017plb}%
  \BibitemOpen
  \bibfield  {author} {\bibinfo {author} {\bibfnamefont {Y.~M.}\ \bibnamefont
  {Gledenov}}, \bibinfo {author} {\bibfnamefont {V.~V.}\ \bibnamefont
  {Nesvizhevsky}}, \bibinfo {author} {\bibfnamefont {P.~V.}\ \bibnamefont
  {Sedyshev}}, \bibinfo {author} {\bibfnamefont {E.~V.}\ \bibnamefont
  {Shul'gina}}, \bibinfo {author} {\bibfnamefont {P.}~\bibnamefont
  {Szalanski}}, \ and\ \bibinfo {author} {\bibfnamefont {V.~A.}\ \bibnamefont
  {Vesna}},\ }\bibfield  {title} {\enquote {\bibinfo {title} {First observation
  of {P}-odd asymmetry of {$\alpha$} particle emission in the
  {$^{10}B(n,\alpha)^7Li$} nuclear reaction},}\ }\href@noop {} {\bibfield
  {journal} {\bibinfo  {journal} {Phys. Lett. B}\ }\textbf {\bibinfo {volume}
  {769}},\ \bibinfo {pages} {111} (\bibinfo {year} {2017})}\BibitemShut
  {NoStop}%
\bibitem [{\citenamefont {Goennenwein}\ \emph {et~al.}(2007)\citenamefont
  {Goennenwein}, \citenamefont {Mutterer}, \citenamefont {Gagarski},
  \citenamefont {Guseva}, \citenamefont {Petrov}, \citenamefont {Sokolov},
  \citenamefont {Zavarukhina}, \citenamefont {Gusev}, \citenamefont {von
  Kalben}, \citenamefont {Nesvizhevsky},\ and\ \citenamefont
  {Soldner}}]{Goe2007plb}%
  \BibitemOpen
  \bibfield  {author} {\bibinfo {author} {\bibfnamefont {F.}~\bibnamefont
  {Goennenwein}}, \bibinfo {author} {\bibfnamefont {M.}~\bibnamefont
  {Mutterer}}, \bibinfo {author} {\bibfnamefont {A.}~\bibnamefont {Gagarski}},
  \bibinfo {author} {\bibfnamefont {I.}~\bibnamefont {Guseva}}, \bibinfo
  {author} {\bibfnamefont {G.}~\bibnamefont {Petrov}}, \bibinfo {author}
  {\bibfnamefont {V.}~\bibnamefont {Sokolov}}, \bibinfo {author} {\bibfnamefont
  {T.}~\bibnamefont {Zavarukhina}}, \bibinfo {author} {\bibfnamefont
  {Y.}~\bibnamefont {Gusev}}, \bibinfo {author} {\bibfnamefont
  {J.}~\bibnamefont {von Kalben}}, \bibinfo {author} {\bibfnamefont {V.~V.}\
  \bibnamefont {Nesvizhevsky}}, \ and\ \bibinfo {author} {\bibfnamefont
  {T.}~\bibnamefont {Soldner}},\ }\bibfield  {title} {\enquote {\bibinfo
  {title} {Rotation of the compound nucleus {$^{236}{\rm U}^*$} in the fission
  reaction {$^{235}{\rm U}(n, f)$} induced by cold polarised neutrons},}\
  }\href@noop {} {\bibfield  {journal} {\bibinfo  {journal} {Phys. Lett. B}\
  }\textbf {\bibinfo {volume} {652}},\ \bibinfo {pages} {13} (\bibinfo {year}
  {2007})}\BibitemShut {NoStop}%
\bibitem [{\citenamefont {Gagarski}\ \emph {et~al.}(2016)\citenamefont
  {Gagarski}, \citenamefont {Goennenwein}, \citenamefont {Guseva},
  \citenamefont {Jesinger}, \citenamefont {Kopatch}, \citenamefont {Kuzmina},
  \citenamefont {Lelievre-Berna}, \citenamefont {Mutterer}, \citenamefont
  {Nesvizhevsky}, \citenamefont {Petrov}, \citenamefont {Soldner},
  \citenamefont {Tiourine}, \citenamefont {Trzaska},\ and\ \citenamefont
  {Zavarukhina}}]{Gag2016prc}%
  \BibitemOpen
  \bibfield  {author} {\bibinfo {author} {\bibfnamefont {A.}~\bibnamefont
  {Gagarski}}, \bibinfo {author} {\bibfnamefont {F.}~\bibnamefont
  {Goennenwein}}, \bibinfo {author} {\bibfnamefont {I.}~\bibnamefont {Guseva}},
  \bibinfo {author} {\bibfnamefont {P.}~\bibnamefont {Jesinger}}, \bibinfo
  {author} {\bibfnamefont {Y.}~\bibnamefont {Kopatch}}, \bibinfo {author}
  {\bibfnamefont {T.}~\bibnamefont {Kuzmina}}, \bibinfo {author} {\bibfnamefont
  {E.}~\bibnamefont {Lelievre-Berna}}, \bibinfo {author} {\bibfnamefont
  {M.}~\bibnamefont {Mutterer}}, \bibinfo {author} {\bibfnamefont {V.~V.}\
  \bibnamefont {Nesvizhevsky}}, \bibinfo {author} {\bibfnamefont
  {G.}~\bibnamefont {Petrov}}, \bibinfo {author} {\bibfnamefont
  {T.}~\bibnamefont {Soldner}}, \bibinfo {author} {\bibfnamefont
  {G.}~\bibnamefont {Tiourine}}, \bibinfo {author} {\bibfnamefont
  {W.}~\bibnamefont {Trzaska}}, \ and\ \bibinfo {author} {\bibfnamefont
  {T.}~\bibnamefont {Zavarukhina}},\ }\bibfield  {title} {\enquote {\bibinfo
  {title} {Particular features of ternary fission induced by polarized neutrons
  in the major actinides {$^{233,235}$U} and {$^{239,241}$Pu}},}\ }\href@noop
  {} {\bibfield  {journal} {\bibinfo  {journal} {Phys. Rev. C}\ }\textbf
  {\bibinfo {volume} {93}},\ \bibinfo {pages} {054619} (\bibinfo {year}
  {2016})}\BibitemShut {NoStop}%
\bibitem [{\citenamefont {M{\"a}rkisch}\ \emph {et~al.}(2019)\citenamefont
  {M{\"a}rkisch}, \citenamefont {Mest}, \citenamefont {Saul}, \citenamefont
  {Wang}, \citenamefont {Abele}, \citenamefont {Dubbers}, \citenamefont
  {Klopf}, \citenamefont {Petukhov}, \citenamefont {Roick}, \citenamefont
  {Soldner},\ and\ \citenamefont {Werder}}]{Mar2019prl}%
  \BibitemOpen
  \bibfield  {author} {\bibinfo {author} {\bibfnamefont {B.}~\bibnamefont
  {M{\"a}rkisch}}, \bibinfo {author} {\bibfnamefont {H.}~\bibnamefont {Mest}},
  \bibinfo {author} {\bibfnamefont {H.}~\bibnamefont {Saul}}, \bibinfo {author}
  {\bibfnamefont {X.}~\bibnamefont {Wang}}, \bibinfo {author} {\bibfnamefont
  {H.}~\bibnamefont {Abele}}, \bibinfo {author} {\bibfnamefont
  {D.}~\bibnamefont {Dubbers}}, \bibinfo {author} {\bibfnamefont
  {M.}~\bibnamefont {Klopf}}, \bibinfo {author} {\bibfnamefont
  {A.}~\bibnamefont {Petukhov}}, \bibinfo {author} {\bibfnamefont
  {C.}~\bibnamefont {Roick}}, \bibinfo {author} {\bibfnamefont
  {T.}~\bibnamefont {Soldner}}, \ and\ \bibinfo {author} {\bibfnamefont
  {D.}~\bibnamefont {Werder}},\ }\bibfield  {title} {\enquote {\bibinfo {title}
  {Measurement of the weak axial-vector coupling constant in the decay of free
  neutrons using a pulsed cold neutron beam},}\ }\href@noop {} {\bibfield
  {journal} {\bibinfo  {journal} {Phys. Rev. Lett.}\ }\textbf {\bibinfo
  {volume} {122}},\ \bibinfo {pages} {242501} (\bibinfo {year}
  {2019})}\BibitemShut {NoStop}%
\bibitem [{\citenamefont {Kreuz}\ \emph
  {et~al.}(2005{\natexlab{a}})\citenamefont {Kreuz}, \citenamefont {Soldner},
  \citenamefont {Baessler}, \citenamefont {Brand}, \citenamefont {Gluck},
  \citenamefont {Mayer}, \citenamefont {Mund}, \citenamefont {Nesvizhevsky},
  \citenamefont {Petoukhov}, \citenamefont {Plonka}, \citenamefont {Reich},
  \citenamefont {Vogel},\ and\ \citenamefont {Abele}}]{Kre2005plb}%
  \BibitemOpen
  \bibfield  {author} {\bibinfo {author} {\bibfnamefont {M.}~\bibnamefont
  {Kreuz}}, \bibinfo {author} {\bibfnamefont {T.}~\bibnamefont {Soldner}},
  \bibinfo {author} {\bibfnamefont {S.}~\bibnamefont {Baessler}}, \bibinfo
  {author} {\bibfnamefont {B.}~\bibnamefont {Brand}}, \bibinfo {author}
  {\bibfnamefont {F.}~\bibnamefont {Gluck}}, \bibinfo {author} {\bibfnamefont
  {U.}~\bibnamefont {Mayer}}, \bibinfo {author} {\bibfnamefont
  {D.}~\bibnamefont {Mund}}, \bibinfo {author} {\bibfnamefont {V.~V.}\
  \bibnamefont {Nesvizhevsky}}, \bibinfo {author} {\bibfnamefont
  {A.}~\bibnamefont {Petoukhov}}, \bibinfo {author} {\bibfnamefont
  {C.}~\bibnamefont {Plonka}}, \bibinfo {author} {\bibfnamefont
  {J.}~\bibnamefont {Reich}}, \bibinfo {author} {\bibfnamefont
  {C.}~\bibnamefont {Vogel}}, \ and\ \bibinfo {author} {\bibfnamefont
  {H.}~\bibnamefont {Abele}},\ }\bibfield  {title} {\enquote {\bibinfo {title}
  {A measurement of the antineutrino asymmetry {B} in the free neutron
  decay},}\ }\href@noop {} {\bibfield  {journal} {\bibinfo  {journal} {Phys.
  Lett. B}\ }\textbf {\bibinfo {volume} {619}},\ \bibinfo {pages} {263}
  (\bibinfo {year} {2005}{\natexlab{a}})}\BibitemShut {NoStop}%
\bibitem [{\citenamefont {Mezei}\ and\ \citenamefont
  {Dagleish}(1977)}]{Mez1977cp}%
  \BibitemOpen
  \bibfield  {author} {\bibinfo {author} {\bibfnamefont {F.}~\bibnamefont
  {Mezei}}\ and\ \bibinfo {author} {\bibfnamefont {P.~A.}\ \bibnamefont
  {Dagleish}},\ }\bibfield  {title} {\enquote {\bibinfo {title} {Corrigendum
  and 1st experimental evidence on neutron supermirrors},}\ }\href@noop {}
  {\bibfield  {journal} {\bibinfo  {journal} {Commun. Phys.}\ }\textbf
  {\bibinfo {volume} {2}},\ \bibinfo {pages} {41} (\bibinfo {year}
  {1977})}\BibitemShut {NoStop}%
\bibitem [{\citenamefont {Drabkin}\ \emph {et~al.}(1977)\citenamefont
  {Drabkin}, \citenamefont {Okorokov}, \citenamefont {Shchebetov},
  \citenamefont {Borovikova}, \citenamefont {Gukasov}, \citenamefont {Korneev},
  \citenamefont {Kudryashov},\ and\ \citenamefont {Runov}}]{Dra1977jtp}%
  \BibitemOpen
  \bibfield  {author} {\bibinfo {author} {\bibfnamefont {G.~M.}\ \bibnamefont
  {Drabkin}}, \bibinfo {author} {\bibfnamefont {A.}~\bibnamefont {Okorokov}},
  \bibinfo {author} {\bibfnamefont {A.~F.}\ \bibnamefont {Shchebetov}},
  \bibinfo {author} {\bibfnamefont {N.~V.}\ \bibnamefont {Borovikova}},
  \bibinfo {author} {\bibfnamefont {A.~G.}\ \bibnamefont {Gukasov}}, \bibinfo
  {author} {\bibfnamefont {D.~A.}\ \bibnamefont {Korneev}}, \bibinfo {author}
  {\bibfnamefont {V.~A.}\ \bibnamefont {Kudryashov}}, \ and\ \bibinfo {author}
  {\bibfnamefont {V.~V.}\ \bibnamefont {Runov}},\ }\bibfield  {title} {\enquote
  {\bibinfo {title} {Polarizing neutron guide on basis of multilayer
  mirrors},}\ }\href@noop {} {\bibfield  {journal} {\bibinfo  {journal} {J.
  Tech. Phys.}\ }\textbf {\bibinfo {volume} {47}},\ \bibinfo {pages} {203}
  (\bibinfo {year} {1977})}\BibitemShut {NoStop}%
\bibitem [{\citenamefont {Schaerpf}(1989)}]{Sch1989pb}%
  \BibitemOpen
  \bibfield  {author} {\bibinfo {author} {\bibfnamefont {O.}~\bibnamefont
  {Schaerpf}},\ }\bibfield  {title} {\enquote {\bibinfo {title} {Properties of
  beam bender type neutron polarizers using supermirrors},}\ }\href@noop {}
  {\bibfield  {journal} {\bibinfo  {journal} {Phys. B}\ }\textbf {\bibinfo
  {volume} {156}},\ \bibinfo {pages} {639} (\bibinfo {year}
  {1989})}\BibitemShut {NoStop}%
\bibitem [{\citenamefont {Majkrzak}(1995)}]{Maj1995pb}%
  \BibitemOpen
  \bibfield  {author} {\bibinfo {author} {\bibfnamefont {C.~F.}\ \bibnamefont
  {Majkrzak}},\ }\bibfield  {title} {\enquote {\bibinfo {title} {Advances in
  polarized neutrons reflectometry},}\ }\href@noop {} {\bibfield  {journal}
  {\bibinfo  {journal} {Phys. B}\ }\textbf {\bibinfo {volume} {213}},\ \bibinfo
  {pages} {904} (\bibinfo {year} {1995})}\BibitemShut {NoStop}%
\bibitem [{\citenamefont {Maruyama}\ \emph {et~al.}(2007)\citenamefont
  {Maruyama}, \citenamefont {Yamazaki}, \citenamefont {Ebisawa}, \citenamefont
  {Hino},\ and\ \citenamefont {Soyama}}]{Mar2007tsf}%
  \BibitemOpen
  \bibfield  {author} {\bibinfo {author} {\bibfnamefont {R.}~\bibnamefont
  {Maruyama}}, \bibinfo {author} {\bibfnamefont {T.}~\bibnamefont {Yamazaki}},
  \bibinfo {author} {\bibfnamefont {T.}~\bibnamefont {Ebisawa}}, \bibinfo
  {author} {\bibfnamefont {M.}~\bibnamefont {Hino}}, \ and\ \bibinfo {author}
  {\bibfnamefont {K.}~\bibnamefont {Soyama}},\ }\bibfield  {title} {\enquote
  {\bibinfo {title} {Development of neutron supermirrors with large critical
  angle},}\ }\href@noop {} {\bibfield  {journal} {\bibinfo  {journal} {Thin
  Sol. Film}\ }\textbf {\bibinfo {volume} {515}},\ \bibinfo {pages} {5704}
  (\bibinfo {year} {2007})}\BibitemShut {NoStop}%
\bibitem [{\citenamefont {Krist}\ \emph {et~al.}(2008)\citenamefont {Krist},
  \citenamefont {Teichert}, \citenamefont {Mezei},\ and\ \citenamefont
  {Rosta}}]{Kri2008}%
  \BibitemOpen
  \bibfield  {author} {\bibinfo {author} {\bibfnamefont {T.}~\bibnamefont
  {Krist}}, \bibinfo {author} {\bibfnamefont {A.}~\bibnamefont {Teichert}},
  \bibinfo {author} {\bibfnamefont {F.}~\bibnamefont {Mezei}}, \ and\ \bibinfo
  {author} {\bibfnamefont {L.}~\bibnamefont {Rosta}},\ }\href@noop {} {\emph
  {\bibinfo {title} {Modern Developments in {X}-Ray and Neutron Optics, pp.
  355-370}}}\ (\bibinfo  {publisher} {Springer-Verlag, Berlin, Heidelberg,
  ISBN: 978-3-540-74560-0},\ \bibinfo {year} {2008})\BibitemShut {NoStop}%
\bibitem [{\citenamefont {Mezei}(1989)}]{Mez1989spie}%
  \BibitemOpen
  \bibfield  {author} {\bibinfo {author} {\bibfnamefont {F.}~\bibnamefont
  {Mezei}},\ }\bibfield  {title} {\enquote {\bibinfo {title} {Very high
  reflectivity supermirrors and their applications},}\ }\href@noop {}
  {\bibfield  {journal} {\bibinfo  {journal} {Proc. SPIE}\ }\textbf {\bibinfo
  {volume} {983}},\ \bibinfo {pages} {10} (\bibinfo {year} {1989})}\BibitemShut
  {NoStop}%
\bibitem [{\citenamefont {Pleshanov}(2010)}]{Ple2010nima}%
  \BibitemOpen
  \bibfield  {author} {\bibinfo {author} {\bibfnamefont {N.~K.}\ \bibnamefont
  {Pleshanov}},\ }\bibfield  {title} {\enquote {\bibinfo {title}
  {Superpolarizing neutron coatings: {T}heory and first experiments},}\
  }\href@noop {} {\bibfield  {journal} {\bibinfo  {journal} {Nucl. Instr. Meth.
  A}\ }\textbf {\bibinfo {volume} {613}},\ \bibinfo {pages} {12} (\bibinfo
  {year} {2010})}\BibitemShut {NoStop}%
\bibitem [{\citenamefont {Kreuz}\ \emph
  {et~al.}(2005{\natexlab{b}})\citenamefont {Kreuz}, \citenamefont
  {Nesvizhevsky}, \citenamefont {Petukhov},\ and\ \citenamefont
  {Soldner}}]{Kre2005nima}%
  \BibitemOpen
  \bibfield  {author} {\bibinfo {author} {\bibfnamefont {M.}~\bibnamefont
  {Kreuz}}, \bibinfo {author} {\bibfnamefont {V.~V.}\ \bibnamefont
  {Nesvizhevsky}}, \bibinfo {author} {\bibfnamefont {A.}~\bibnamefont
  {Petukhov}}, \ and\ \bibinfo {author} {\bibfnamefont {T.}~\bibnamefont
  {Soldner}},\ }\bibfield  {title} {\enquote {\bibinfo {title} {The crossed
  geometry of two supermirror polarizers - a new method for neutron beam
  polarization and polarization analysis},}\ }\href@noop {} {\bibfield
  {journal} {\bibinfo  {journal} {Nucl. Instr. Meth. A}\ }\textbf {\bibinfo
  {volume} {547}},\ \bibinfo {pages} {583} (\bibinfo {year}
  {2005}{\natexlab{b}})}\BibitemShut {NoStop}%
\bibitem [{\citenamefont {Petukhov}\ \emph {et~al.}(2016)\citenamefont
  {Petukhov}, \citenamefont {Nesvizhevsky}, \citenamefont {Bigault},
  \citenamefont {Courtois}, \citenamefont {Jullien},\ and\ \citenamefont
  {Soldner}}]{Pet2016nima}%
  \BibitemOpen
  \bibfield  {author} {\bibinfo {author} {\bibfnamefont {A.~K.}\ \bibnamefont
  {Petukhov}}, \bibinfo {author} {\bibfnamefont {V.~V.}\ \bibnamefont
  {Nesvizhevsky}}, \bibinfo {author} {\bibfnamefont {T.}~\bibnamefont
  {Bigault}}, \bibinfo {author} {\bibfnamefont {P.}~\bibnamefont {Courtois}},
  \bibinfo {author} {\bibfnamefont {D.}~\bibnamefont {Jullien}}, \ and\
  \bibinfo {author} {\bibfnamefont {T.}~\bibnamefont {Soldner}},\ }\bibfield
  {title} {\enquote {\bibinfo {title} {A concept of advanced broad-band
  solid-state supermirror polarizers for cold neutrons},}\ }\href@noop {}
  {\bibfield  {journal} {\bibinfo  {journal} {Nucl. Instr. Meth. A}\ }\textbf
  {\bibinfo {volume} {838}},\ \bibinfo {pages} {33} (\bibinfo {year}
  {2016})}\BibitemShut {NoStop}%
\bibitem [{\citenamefont {Elscnhans}\ \emph {et~al.}(1994)\citenamefont
  {Elscnhans}, \citenamefont {Boni}, \citenamefont {Fricdli}, \citenamefont
  {Grimmer}, \citenamefont {Buffat}, \citenamefont {Lcifcr}, \citenamefont
  {Sochug},\ and\ \citenamefont {Anderson}}]{Els1994tsf}%
  \BibitemOpen
  \bibfield  {author} {\bibinfo {author} {\bibfnamefont {O.}~\bibnamefont
  {Elscnhans}}, \bibinfo {author} {\bibfnamefont {P.}~\bibnamefont {Boni}},
  \bibinfo {author} {\bibfnamefont {H.~P.}\ \bibnamefont {Fricdli}}, \bibinfo
  {author} {\bibfnamefont {H.}~\bibnamefont {Grimmer}}, \bibinfo {author}
  {\bibfnamefont {P.}~\bibnamefont {Buffat}}, \bibinfo {author} {\bibfnamefont
  {K.}~\bibnamefont {Lcifcr}}, \bibinfo {author} {\bibfnamefont
  {J.}~\bibnamefont {Sochug}}, \ and\ \bibinfo {author} {\bibfnamefont {I.~A.}\
  \bibnamefont {Anderson}},\ }\bibfield  {title} {\enquote {\bibinfo {title}
  {Development of {N}i {T}i multilayer supermirrors for neutron optics},}\
  }\href@noop {} {\bibfield  {journal} {\bibinfo  {journal} {Thin Sol. Films}\
  }\textbf {\bibinfo {volume} {246}},\ \bibinfo {pages} {110} (\bibinfo {year}
  {1994})}\BibitemShut {NoStop}%
\bibitem [{\citenamefont {Courtois}\ \emph {et~al.}(2013)\citenamefont
  {Courtois}, \citenamefont {Bigault}, \citenamefont {Gaignon}, \citenamefont
  {Vottoz}, \citenamefont {Delphin},\ and\ \citenamefont
  {Bourgault}}]{Cou2013ILL}%
  \BibitemOpen
  \bibfield  {author} {\bibinfo {author} {\bibfnamefont {P.}~\bibnamefont
  {Courtois}}, \bibinfo {author} {\bibfnamefont {T.}~\bibnamefont {Bigault}},
  \bibinfo {author} {\bibfnamefont {V.}~\bibnamefont {Gaignon}}, \bibinfo
  {author} {\bibfnamefont {A.}~\bibnamefont {Vottoz}}, \bibinfo {author}
  {\bibfnamefont {G.}~\bibnamefont {Delphin}}, \ and\ \bibinfo {author}
  {\bibfnamefont {D.}~\bibnamefont {Bourgault}},\ }\bibfield  {title} {\enquote
  {\bibinfo {title} {Co/{T}i supermirror analyzer with large detector area
  coverage for {WASP}},}\ }in\ \href@noop {} {\emph {\bibinfo {booktitle} {ILL
  Annual Report 2013}}}\ (\bibinfo {organization} {Institut Laue Langevin},\
  \bibinfo {year} {2013})\ p.~\bibinfo {pages} {82}\BibitemShut {NoStop}%
\bibitem [{\citenamefont {Soldner}, \citenamefont {Petukhov},\ and\
  \citenamefont {Plonka}(2002)}]{Sol2002ILL}%
  \BibitemOpen
  \bibfield  {author} {\bibinfo {author} {\bibfnamefont {T.}~\bibnamefont
  {Soldner}}, \bibinfo {author} {\bibfnamefont {A.}~\bibnamefont {Petukhov}}, \
  and\ \bibinfo {author} {\bibfnamefont {C.}~\bibnamefont {Plonka}},\
  }\href@noop {} {\enquote {\bibinfo {title} {Installation and first tests of
  the new {PF1b} polariser},}\ }\bibinfo {type} {Tech. Rep.}\ \bibinfo {number}
  {ILL Technical Report ILL03{S}010{T}}\ (\bibinfo  {institution} {Institut
  Laue Langevin},\ \bibinfo {year} {2002})\BibitemShut {NoStop}%
\bibitem [{\citenamefont {Petukhov}\ \emph {et~al.}(2019)\citenamefont
  {Petukhov}, \citenamefont {Nesvizhevsky}, \citenamefont {Bigault},
  \citenamefont {Courtois}, \citenamefont {Jullien},\ and\ \citenamefont
  {Soldner}}]{Pet2019rsi}%
  \BibitemOpen
  \bibfield  {author} {\bibinfo {author} {\bibfnamefont {A.~K.}\ \bibnamefont
  {Petukhov}}, \bibinfo {author} {\bibfnamefont {V.~V.}\ \bibnamefont
  {Nesvizhevsky}}, \bibinfo {author} {\bibfnamefont {T.}~\bibnamefont
  {Bigault}}, \bibinfo {author} {\bibfnamefont {P.}~\bibnamefont {Courtois}},
  \bibinfo {author} {\bibfnamefont {D.}~\bibnamefont {Jullien}}, \ and\
  \bibinfo {author} {\bibfnamefont {T.}~\bibnamefont {Soldner}},\ }\bibfield
  {title} {\enquote {\bibinfo {title} {A project of advanced solid-state
  neutron polarizer for {PF}1{B} instrument at {I}nstitut {L}aue-{L}angevin},}\
  }\href@noop {} {\bibfield  {journal} {\bibinfo  {journal} {Rev. Sci. Instr.}\
  }\textbf {\bibinfo {volume} {547}},\ \bibinfo {pages} {583} (\bibinfo {year}
  {2019})}\BibitemShut {NoStop}%
\bibitem [{\citenamefont {Krist}\ \emph {et~al.}(1998)\citenamefont {Krist},
  \citenamefont {Kennedy}, \citenamefont {Hicks},\ and\ \citenamefont
  {Mezei}}]{Kri1998pb}%
  \BibitemOpen
  \bibfield  {author} {\bibinfo {author} {\bibfnamefont {T.}~\bibnamefont
  {Krist}}, \bibinfo {author} {\bibfnamefont {S.~J.}\ \bibnamefont {Kennedy}},
  \bibinfo {author} {\bibfnamefont {T.}~\bibnamefont {Hicks}}, \ and\ \bibinfo
  {author} {\bibfnamefont {F.}~\bibnamefont {Mezei}},\ }\bibfield  {title}
  {\enquote {\bibinfo {title} {New compact neutron polarizer},}\ }\href@noop {}
  {\bibfield  {journal} {\bibinfo  {journal} {Phys. B}\ }\textbf {\bibinfo
  {volume} {241}},\ \bibinfo {pages} {82} (\bibinfo {year} {1998})}\BibitemShut
  {NoStop}%
\bibitem [{\citenamefont {Hoghoj}\ \emph {et~al.}(1999)\citenamefont {Hoghoj},
  \citenamefont {Anderson}, \citenamefont {Siebrecht}, \citenamefont {Graf},\
  and\ \citenamefont {Ben-Sadidane}}]{Hog1999pb}%
  \BibitemOpen
  \bibfield  {author} {\bibinfo {author} {\bibfnamefont {P.}~\bibnamefont
  {Hoghoj}}, \bibinfo {author} {\bibfnamefont {I.}~\bibnamefont {Anderson}},
  \bibinfo {author} {\bibfnamefont {R.}~\bibnamefont {Siebrecht}}, \bibinfo
  {author} {\bibfnamefont {W.}~\bibnamefont {Graf}}, \ and\ \bibinfo {author}
  {\bibfnamefont {K.}~\bibnamefont {Ben-Sadidane}},\ }\bibfield  {title}
  {\enquote {\bibinfo {title} {Neutron polarizing {F}e{S}i mirrors at {ILL}},}\
  }\href@noop {} {\bibfield  {journal} {\bibinfo  {journal} {Phys. B}\ }\textbf
  {\bibinfo {volume} {267}},\ \bibinfo {pages} {355} (\bibinfo {year}
  {1999})}\BibitemShut {NoStop}%
\bibitem [{\citenamefont {Stunault}\ \emph {et~al.}(2006)\citenamefont
  {Stunault}, \citenamefont {Andersen}, \citenamefont {Roux}, \citenamefont
  {Bigault}, \citenamefont {Ben-Saidane},\ and\ \citenamefont
  {Ronnow}}]{Stu2006pb}%
  \BibitemOpen
  \bibfield  {author} {\bibinfo {author} {\bibfnamefont {A.}~\bibnamefont
  {Stunault}}, \bibinfo {author} {\bibfnamefont {K.~H.}\ \bibnamefont
  {Andersen}}, \bibinfo {author} {\bibfnamefont {S.}~\bibnamefont {Roux}},
  \bibinfo {author} {\bibfnamefont {T.}~\bibnamefont {Bigault}}, \bibinfo
  {author} {\bibfnamefont {K.}~\bibnamefont {Ben-Saidane}}, \ and\ \bibinfo
  {author} {\bibfnamefont {H.~M.}\ \bibnamefont {Ronnow}},\ }\bibfield  {title}
  {\enquote {\bibinfo {title} {Magnetic excitations in a new anysotropic kagome
  antiferromagnetic},}\ }\href@noop {} {\bibfield  {journal} {\bibinfo
  {journal} {Phys. B}\ }\textbf {\bibinfo {volume} {385}},\ \bibinfo {pages}
  {1152} (\bibinfo {year} {2006})}\BibitemShut {NoStop}%
\bibitem [{\citenamefont {Bigault}\ \emph {et~al.}(2009)\citenamefont
  {Bigault}, \citenamefont {Andersen}, \citenamefont {Hiess}, \citenamefont
  {Roux}, \citenamefont {Stunault}, \citenamefont {Bisig},\ and\ \citenamefont
  {Bouvier}}]{Big2009ILL}%
  \BibitemOpen
  \bibfield  {author} {\bibinfo {author} {\bibfnamefont {T.}~\bibnamefont
  {Bigault}}, \bibinfo {author} {\bibfnamefont {K.~H.}\ \bibnamefont
  {Andersen}}, \bibinfo {author} {\bibfnamefont {A.}~\bibnamefont {Hiess}},
  \bibinfo {author} {\bibfnamefont {S.}~\bibnamefont {Roux}}, \bibinfo {author}
  {\bibfnamefont {A.}~\bibnamefont {Stunault}}, \bibinfo {author}
  {\bibfnamefont {G.}~\bibnamefont {Bisig}}, \ and\ \bibinfo {author}
  {\bibfnamefont {A.}~\bibnamefont {Bouvier}},\ }\bibfield  {title} {\enquote
  {\bibinfo {title} {Hey, big {S}-bender},}\ }in\ \href@noop {} {\emph
  {\bibinfo {booktitle} {ILL Annual Report 2009}}}\ (\bibinfo {organization}
  {Institut Laue Langevin},\ \bibinfo {year} {2009})\ p.~\bibinfo {pages}
  {94}\BibitemShut {NoStop}%
\bibitem [{\citenamefont {Wildes}()}]{Wil2011itr}%
  \BibitemOpen
  \bibfield  {author} {\bibinfo {author} {\bibfnamefont {A.~R.}\ \bibnamefont
  {Wildes}},\ }\bibfield  {title} {\enquote {\bibinfo {title} {On the
  performance of the {D}17 {S}-bender},}\ }\href@noop {} {\bibfield  {journal}
  {\bibinfo  {journal} {Int. Tech. Rep. ILL}\ }\textbf {\bibinfo {volume}
  {2011}}}\BibitemShut {NoStop}%
\bibitem [{\citenamefont {Shah}\ \emph {et~al.}(2014)\citenamefont {Shah},
  \citenamefont {Washington}, \citenamefont {Stonaha}, \citenamefont {Ashkar},
  \citenamefont {Kaiser}, \citenamefont {Krist},\ and\ \citenamefont
  {Pynn}}]{Sha2014nima}%
  \BibitemOpen
  \bibfield  {author} {\bibinfo {author} {\bibfnamefont {V.~R.}\ \bibnamefont
  {Shah}}, \bibinfo {author} {\bibfnamefont {A.~L.}\ \bibnamefont
  {Washington}}, \bibinfo {author} {\bibfnamefont {P.}~\bibnamefont {Stonaha}},
  \bibinfo {author} {\bibfnamefont {R.}~\bibnamefont {Ashkar}}, \bibinfo
  {author} {\bibfnamefont {H.}~\bibnamefont {Kaiser}}, \bibinfo {author}
  {\bibfnamefont {T.}~\bibnamefont {Krist}}, \ and\ \bibinfo {author}
  {\bibfnamefont {R.}~\bibnamefont {Pynn}},\ }\bibfield  {title} {\enquote
  {\bibinfo {title} {Optimization of a solid-state polarizing bender for cold
  neutrons},}\ }\href@noop {} {\bibfield  {journal} {\bibinfo  {journal} {Nucl.
  Instr. Meth. A}\ }\textbf {\bibinfo {volume} {768}},\ \bibinfo {pages} {157}
  (\bibinfo {year} {2014})}\BibitemShut {NoStop}%
\bibitem [{\citenamefont {Katyba}\ \emph {et~al.}(2018)\citenamefont {Katyba},
  \citenamefont {Zaytsev}, \citenamefont {Dolganova}, \citenamefont
  {Shikunova}, \citenamefont {Chernomyrdin}, \citenamefont {Yurchenko},
  \citenamefont {Komandin}, \citenamefont {Reshetov}, \citenamefont
  {Nesvizhevsky},\ and\ \citenamefont {Kurlov}}]{Kat2018pcgcm}%
  \BibitemOpen
  \bibfield  {author} {\bibinfo {author} {\bibfnamefont {G.~M.}\ \bibnamefont
  {Katyba}}, \bibinfo {author} {\bibfnamefont {K.~I.}\ \bibnamefont {Zaytsev}},
  \bibinfo {author} {\bibfnamefont {I.~N.}\ \bibnamefont {Dolganova}}, \bibinfo
  {author} {\bibfnamefont {A.~I.}\ \bibnamefont {Shikunova}}, \bibinfo {author}
  {\bibfnamefont {N.~V.}\ \bibnamefont {Chernomyrdin}}, \bibinfo {author}
  {\bibfnamefont {S.~O.}\ \bibnamefont {Yurchenko}}, \bibinfo {author}
  {\bibfnamefont {G.~A.}\ \bibnamefont {Komandin}}, \bibinfo {author}
  {\bibfnamefont {I.~V.}\ \bibnamefont {Reshetov}}, \bibinfo {author}
  {\bibfnamefont {V.~V.}\ \bibnamefont {Nesvizhevsky}}, \ and\ \bibinfo
  {author} {\bibfnamefont {V.~N.}\ \bibnamefont {Kurlov}},\ }\bibfield  {title}
  {\enquote {\bibinfo {title} {Sapphire shaped crystals for waveguiding,
  sensing and exposure applications},}\ }\href@noop {} {\bibfield  {journal}
  {\bibinfo  {journal} {Progr. Cryst. Grow. Charact. Mater.}\ }\textbf
  {\bibinfo {volume} {64}},\ \bibinfo {pages} {133} (\bibinfo {year}
  {2018})}\BibitemShut {NoStop}%
\bibitem [{Sie()}]{Siegert}%
  \BibitemOpen
  \href@noop {} {}\bibinfo {note} {Siegert Wafer GmbH, Charlottenburger Allee
  7, 52068 Aachen, Germany, http://www.siegertwafer.com}\BibitemShut {NoStop}%
\bibitem [{\citenamefont {Scharpf}\ and\ \citenamefont
  {Anderson}(1994)}]{Sch1994pb}%
  \BibitemOpen
  \bibfield  {author} {\bibinfo {author} {\bibfnamefont {O.}~\bibnamefont
  {Scharpf}}\ and\ \bibinfo {author} {\bibfnamefont {I.~S.}\ \bibnamefont
  {Anderson}},\ }\bibfield  {title} {\enquote {\bibinfo {title} {The role of
  surfaces and interfaces in the behavior of non-polarizing and polarizing
  supermirrors},}\ }\href@noop {} {\bibfield  {journal} {\bibinfo  {journal}
  {Phys. B}\ }\textbf {\bibinfo {volume} {198}},\ \bibinfo {pages} {203}
  (\bibinfo {year} {1994})}\BibitemShut {NoStop}%
\bibitem [{\citenamefont {Bigault}\ \emph {et~al.}(2014)\citenamefont
  {Bigault}, \citenamefont {Delphin}, \citenamefont {Vottoz}, \citenamefont
  {Gaignon},\ and\ \citenamefont {Courtois}}]{Big2014jpcs}%
  \BibitemOpen
  \bibfield  {author} {\bibinfo {author} {\bibfnamefont {T.}~\bibnamefont
  {Bigault}}, \bibinfo {author} {\bibfnamefont {G.}~\bibnamefont {Delphin}},
  \bibinfo {author} {\bibfnamefont {A.}~\bibnamefont {Vottoz}}, \bibinfo
  {author} {\bibfnamefont {V.}~\bibnamefont {Gaignon}}, \ and\ \bibinfo
  {author} {\bibfnamefont {P.}~\bibnamefont {Courtois}},\ }\bibfield  {title}
  {\enquote {\bibinfo {title} {Recent polarizing supermirror projects at the
  {ILL}},}\ }\href@noop {} {\bibfield  {journal} {\bibinfo  {journal} {J. Phys.
  Conf. Ser.}\ }\textbf {\bibinfo {volume} {528}},\ \bibinfo {pages} {012017}
  (\bibinfo {year} {2014})}\BibitemShut {NoStop}%
\bibitem [{\citenamefont {Maruyama}\ \emph {et~al.}(2016)\citenamefont
  {Maruyama}, \citenamefont {Bigault}, \citenamefont {Wildes}, \citenamefont
  {Dewhurst}, \citenamefont {Soyama},\ and\ \citenamefont
  {Courtois}}]{Mar2016nima}%
  \BibitemOpen
  \bibfield  {author} {\bibinfo {author} {\bibfnamefont {R.}~\bibnamefont
  {Maruyama}}, \bibinfo {author} {\bibfnamefont {T.}~\bibnamefont {Bigault}},
  \bibinfo {author} {\bibfnamefont {A.~R.}\ \bibnamefont {Wildes}}, \bibinfo
  {author} {\bibfnamefont {C.~D.}\ \bibnamefont {Dewhurst}}, \bibinfo {author}
  {\bibfnamefont {K.}~\bibnamefont {Soyama}}, \ and\ \bibinfo {author}
  {\bibfnamefont {P.}~\bibnamefont {Courtois}},\ }\bibfield  {title} {\enquote
  {\bibinfo {title} {Study of the in-plane magnetic structure of a layered
  system using polarized neutron scattering under grazing incidence
  geometry},}\ }\href@noop {} {\bibfield  {journal} {\bibinfo  {journal} {Nucl.
  Instr. Meth. A}\ }\textbf {\bibinfo {volume} {819}},\ \bibinfo {pages} {37}
  (\bibinfo {year} {2016})}\BibitemShut {NoStop}%
\bibitem [{\citenamefont {Maruyama}\ \emph {et~al.}(2019)\citenamefont
  {Maruyama}, \citenamefont {Bigault}, \citenamefont {Saerbeck}, \citenamefont
  {Honecker}, \citenamefont {Soyama},\ and\ \citenamefont
  {Courtois}}]{Mar2019cry}%
  \BibitemOpen
  \bibfield  {author} {\bibinfo {author} {\bibfnamefont {R.}~\bibnamefont
  {Maruyama}}, \bibinfo {author} {\bibfnamefont {T.}~\bibnamefont {Bigault}},
  \bibinfo {author} {\bibfnamefont {T.}~\bibnamefont {Saerbeck}}, \bibinfo
  {author} {\bibfnamefont {D.}~\bibnamefont {Honecker}}, \bibinfo {author}
  {\bibfnamefont {K.}~\bibnamefont {Soyama}}, \ and\ \bibinfo {author}
  {\bibfnamefont {P.}~\bibnamefont {Courtois}},\ }\bibfield  {title} {\enquote
  {\bibinfo {title} {Coherent magnetization rotation of a layered system
  observed by polarized neutron scattering under grazing incidence geometry},}\
  }\href@noop {} {\bibfield  {journal} {\bibinfo  {journal} {Cryst.}\ }\textbf
  {\bibinfo {volume} {9}},\ \bibinfo {pages} {383} (\bibinfo {year}
  {2019})}\BibitemShut {NoStop}%
\bibitem [{\citenamefont {Klauser}\ \emph {et~al.}(2013)\citenamefont
  {Klauser}, \citenamefont {Bigault}, \citenamefont {Rebrova},\ and\
  \citenamefont {Soldner}}]{Kla2013pp}%
  \BibitemOpen
  \bibfield  {author} {\bibinfo {author} {\bibfnamefont {C.}~\bibnamefont
  {Klauser}}, \bibinfo {author} {\bibfnamefont {T.}~\bibnamefont {Bigault}},
  \bibinfo {author} {\bibfnamefont {N.}~\bibnamefont {Rebrova}}, \ and\
  \bibinfo {author} {\bibfnamefont {T.}~\bibnamefont {Soldner}},\ }\bibfield
  {title} {\enquote {\bibinfo {title} {Ultra-sensitive depolarization study of
  polarizing {C}o{T}i supermirrors with the opaque test bench},}\ }\href@noop
  {} {\bibfield  {journal} {\bibinfo  {journal} {Phys. Proc.}\ }\textbf
  {\bibinfo {volume} {42}},\ \bibinfo {pages} {99} (\bibinfo {year}
  {2013})}\BibitemShut {NoStop}%
\bibitem [{\citenamefont {Klauser}\ \emph {et~al.}(2016)\citenamefont
  {Klauser}, \citenamefont {Bigault}, \citenamefont {Boni}, \citenamefont
  {Courtois}, \citenamefont {Devishvili}, \citenamefont {Rebrova},
  \citenamefont {Schneider},\ and\ \citenamefont {Soldner}}]{Kla2016nima}%
  \BibitemOpen
  \bibfield  {author} {\bibinfo {author} {\bibfnamefont {C.}~\bibnamefont
  {Klauser}}, \bibinfo {author} {\bibfnamefont {T.}~\bibnamefont {Bigault}},
  \bibinfo {author} {\bibfnamefont {P.}~\bibnamefont {Boni}}, \bibinfo {author}
  {\bibfnamefont {P.}~\bibnamefont {Courtois}}, \bibinfo {author}
  {\bibfnamefont {A.}~\bibnamefont {Devishvili}}, \bibinfo {author}
  {\bibfnamefont {N.}~\bibnamefont {Rebrova}}, \bibinfo {author} {\bibfnamefont
  {M.}~\bibnamefont {Schneider}}, \ and\ \bibinfo {author} {\bibfnamefont
  {T.}~\bibnamefont {Soldner}},\ }\bibfield  {title} {\enquote {\bibinfo
  {title} {Depolarization in polarizing supermirrors},}\ }\href@noop {}
  {\bibfield  {journal} {\bibinfo  {journal} {Nucl. Instr. Meth. A}\ }\textbf
  {\bibinfo {volume} {840}},\ \bibinfo {pages} {181} (\bibinfo {year}
  {2016})}\BibitemShut {NoStop}%
\bibitem [{\citenamefont {Devishvili}\ \emph {et~al.}(2013)\citenamefont
  {Devishvili}, \citenamefont {Zhernenkov}, \citenamefont {Denisson},
  \citenamefont {Toperverg}, \citenamefont {Wolff}, \citenamefont
  {Hjorvarsson},\ and\ \citenamefont {Zabell}}]{Dev2013rsi}%
  \BibitemOpen
  \bibfield  {author} {\bibinfo {author} {\bibfnamefont {A.}~\bibnamefont
  {Devishvili}}, \bibinfo {author} {\bibfnamefont {K.}~\bibnamefont
  {Zhernenkov}}, \bibinfo {author} {\bibfnamefont {A.~J.~C.}\ \bibnamefont
  {Denisson}}, \bibinfo {author} {\bibfnamefont {B.~P.}\ \bibnamefont
  {Toperverg}}, \bibinfo {author} {\bibfnamefont {M.}~\bibnamefont {Wolff}},
  \bibinfo {author} {\bibfnamefont {B.}~\bibnamefont {Hjorvarsson}}, \ and\
  \bibinfo {author} {\bibfnamefont {H.}~\bibnamefont {Zabell}},\ }\bibfield
  {title} {\enquote {\bibinfo {title} {{SuperADAM}: Upgraded polarized neutron
  reflectometer at the {I}nstitut {L}aue-{L}angevin},}\ }\href@noop {}
  {\bibfield  {journal} {\bibinfo  {journal} {Rev. Sci. Instr.}\ }\textbf
  {\bibinfo {volume} {84}},\ \bibinfo {pages} {025112} (\bibinfo {year}
  {2013})}\BibitemShut {NoStop}%
\bibitem [{Tho()}]{Tho}%
  \BibitemOpen
  \href@noop {} {}\bibinfo {note} {Thorlabs, UV-cured optical adhesives,
  http://www.thorlabs.com/newgrouppage9.cfm?objectgroup id=196}\BibitemShut
  {NoStop}%
\bibitem [{\citenamefont {Hayter}, \citenamefont {Penfold},\ and\ \citenamefont
  {Williams}(1978)}]{Hay1978jpe}%
  \BibitemOpen
  \bibfield  {author} {\bibinfo {author} {\bibfnamefont {J.~B.}\ \bibnamefont
  {Hayter}}, \bibinfo {author} {\bibfnamefont {J.}~\bibnamefont {Penfold}}, \
  and\ \bibinfo {author} {\bibfnamefont {W.}~\bibnamefont {Williams}},\
  }\bibfield  {title} {\enquote {\bibinfo {title} {Compact polarizing {S}oller
  guides for cold neutrons},}\ }\href@noop {} {\bibfield  {journal} {\bibinfo
  {journal} {J. Phys. E}\ }\textbf {\bibinfo {volume} {11}},\ \bibinfo {pages}
  {454} (\bibinfo {year} {1978})}\BibitemShut {NoStop}%
\bibitem [{\citenamefont {Petoukhov}\ \emph
  {et~al.}(2006{\natexlab{a}})\citenamefont {Petoukhov}, \citenamefont
  {Guillard}, \citenamefont {Andersen}, \citenamefont {Bourgeat-Lami},
  \citenamefont {Chung}, \citenamefont {Humblot}, \citenamefont {Julien},
  \citenamefont {Lelievre-Berna}, \citenamefont {Soldner}, \citenamefont
  {Tasset},\ and\ \citenamefont {Thomas}}]{Pet2006nima}%
  \BibitemOpen
  \bibfield  {author} {\bibinfo {author} {\bibfnamefont {A.~K.}\ \bibnamefont
  {Petoukhov}}, \bibinfo {author} {\bibfnamefont {V.}~\bibnamefont {Guillard}},
  \bibinfo {author} {\bibfnamefont {K.~H.}\ \bibnamefont {Andersen}}, \bibinfo
  {author} {\bibfnamefont {E.}~\bibnamefont {Bourgeat-Lami}}, \bibinfo {author}
  {\bibfnamefont {R.}~\bibnamefont {Chung}}, \bibinfo {author} {\bibfnamefont
  {H.}~\bibnamefont {Humblot}}, \bibinfo {author} {\bibfnamefont
  {D.}~\bibnamefont {Julien}}, \bibinfo {author} {\bibfnamefont
  {E.}~\bibnamefont {Lelievre-Berna}}, \bibinfo {author} {\bibfnamefont
  {T.}~\bibnamefont {Soldner}}, \bibinfo {author} {\bibfnamefont
  {F.}~\bibnamefont {Tasset}}, \ and\ \bibinfo {author} {\bibfnamefont
  {M.}~\bibnamefont {Thomas}},\ }\bibfield  {title} {\enquote {\bibinfo {title}
  {Compact magnetostatic cavity for polarized {H}e-3 neutron spin filter
  cells},}\ }\href@noop {} {\bibfield  {journal} {\bibinfo  {journal} {Nucl.
  Instr. Meth. A}\ }\textbf {\bibinfo {volume} {560}},\ \bibinfo {pages} {480}
  (\bibinfo {year} {2006}{\natexlab{a}})}\BibitemShut {NoStop}%
\bibitem [{\citenamefont {Babcock}\ \emph {et~al.}(2007)\citenamefont
  {Babcock}, \citenamefont {Petoukhov}, \citenamefont {Chastagnier},
  \citenamefont {Jullien}, \citenamefont {Lelievre-Berna}, \citenamefont
  {Andersen}, \citenamefont {Georgi}, \citenamefont {Masalovich}, \citenamefont
  {Boag}, \citenamefont {Frost},\ and\ \citenamefont {Parnell}}]{Bab2007pb}%
  \BibitemOpen
  \bibfield  {author} {\bibinfo {author} {\bibfnamefont {E.}~\bibnamefont
  {Babcock}}, \bibinfo {author} {\bibfnamefont {A.}~\bibnamefont {Petoukhov}},
  \bibinfo {author} {\bibfnamefont {J.}~\bibnamefont {Chastagnier}}, \bibinfo
  {author} {\bibfnamefont {D.}~\bibnamefont {Jullien}}, \bibinfo {author}
  {\bibfnamefont {E.}~\bibnamefont {Lelievre-Berna}}, \bibinfo {author}
  {\bibfnamefont {K.~H.}\ \bibnamefont {Andersen}}, \bibinfo {author}
  {\bibfnamefont {R.}~\bibnamefont {Georgi}}, \bibinfo {author} {\bibfnamefont
  {S.}~\bibnamefont {Masalovich}}, \bibinfo {author} {\bibfnamefont
  {S.}~\bibnamefont {Boag}}, \bibinfo {author} {\bibfnamefont {C.~D.}\
  \bibnamefont {Frost}}, \ and\ \bibinfo {author} {\bibfnamefont {S.~R.}\
  \bibnamefont {Parnell}},\ }\bibfield  {title} {\enquote {\bibinfo {title}
  {{AFP} flipper devices: polarized {H}e-3 spin flipper and shorter wavelength
  neutron flipper},}\ }\href@noop {} {\bibfield  {journal} {\bibinfo  {journal}
  {Phys. B}\ }\textbf {\bibinfo {volume} {397}},\ \bibinfo {pages} {172}
  (\bibinfo {year} {2007})}\BibitemShut {NoStop}%
\bibitem [{\citenamefont {Clausen}\ \emph {et~al.}(1997)\citenamefont
  {Clausen}, \citenamefont {McMorrow}, \citenamefont {Lefmann}, \citenamefont
  {Aepplia}, \citenamefont {Mason}, \citenamefont {Schr{\"o}der}, \citenamefont
  {Issikii}, \citenamefont {Nohara},\ and\ \citenamefont {Takagi}}]{Cla1997pb}%
  \BibitemOpen
  \bibfield  {author} {\bibinfo {author} {\bibfnamefont {K.~N.}\ \bibnamefont
  {Clausen}}, \bibinfo {author} {\bibfnamefont {D.~F.}\ \bibnamefont
  {McMorrow}}, \bibinfo {author} {\bibfnamefont {K.}~\bibnamefont {Lefmann}},
  \bibinfo {author} {\bibfnamefont {G.}~\bibnamefont {Aepplia}}, \bibinfo
  {author} {\bibfnamefont {T.~E.}\ \bibnamefont {Mason}}, \bibinfo {author}
  {\bibfnamefont {A.}~\bibnamefont {Schr{\"o}der}}, \bibinfo {author}
  {\bibfnamefont {M.}~\bibnamefont {Issikii}}, \bibinfo {author} {\bibfnamefont
  {M.}~\bibnamefont {Nohara}}, \ and\ \bibinfo {author} {\bibfnamefont
  {H.}~\bibnamefont {Takagi}},\ }\bibfield  {title} {\enquote {\bibinfo {title}
  {The {RITA} spectrometer at {RISO} - design considerations and recent
  results},}\ }\href@noop {} {\bibfield  {journal} {\bibinfo  {journal} {Phys.
  B}\ }\textbf {\bibinfo {volume} {240}},\ \bibinfo {pages} {50} (\bibinfo
  {year} {1997})}\BibitemShut {NoStop}%
\bibitem [{\citenamefont {Alsnielsen}(1967)}]{Als1967nima}%
  \BibitemOpen
  \bibfield  {author} {\bibinfo {author} {\bibfnamefont {J.}~\bibnamefont
  {Alsnielsen}},\ }\bibfield  {title} {\enquote {\bibinfo {title} {Corrections
  in gold foil activation method for determination of neutron beam
  intensity},}\ }\href@noop {} {\bibfield  {journal} {\bibinfo  {journal}
  {Nucl. Instr. Meth. A}\ }\textbf {\bibinfo {volume} {50}},\ \bibinfo {pages}
  {191} (\bibinfo {year} {1967})}\BibitemShut {NoStop}%
\bibitem [{\citenamefont {Andersen}\ \emph {et~al.}(2005)\citenamefont
  {Andersen}, \citenamefont {Chung}, \citenamefont {Guillard}, \citenamefont
  {Humblot}, \citenamefont {Jullien}, \citenamefont {Lelievre-Berna},
  \citenamefont {Petoukhov},\ and\ \citenamefont {Tasset}}]{And2005pb}%
  \BibitemOpen
  \bibfield  {author} {\bibinfo {author} {\bibfnamefont {K.~H.}\ \bibnamefont
  {Andersen}}, \bibinfo {author} {\bibfnamefont {R.}~\bibnamefont {Chung}},
  \bibinfo {author} {\bibfnamefont {V.}~\bibnamefont {Guillard}}, \bibinfo
  {author} {\bibfnamefont {H.}~\bibnamefont {Humblot}}, \bibinfo {author}
  {\bibfnamefont {D.}~\bibnamefont {Jullien}}, \bibinfo {author} {\bibfnamefont
  {E.}~\bibnamefont {Lelievre-Berna}}, \bibinfo {author} {\bibfnamefont
  {A.}~\bibnamefont {Petoukhov}}, \ and\ \bibinfo {author} {\bibfnamefont
  {F.}~\bibnamefont {Tasset}},\ }\bibfield  {title} {\enquote {\bibinfo {title}
  {Towards polarization analysis on a thermal time-of-flight spectrometer},}\
  }\href@noop {} {\bibfield  {journal} {\bibinfo  {journal} {Phys. B}\ }\textbf
  {\bibinfo {volume} {356}},\ \bibinfo {pages} {103} (\bibinfo {year}
  {2005})}\BibitemShut {NoStop}%
\bibitem [{\citenamefont {Petoukhov}\ \emph
  {et~al.}(2006{\natexlab{b}})\citenamefont {Petoukhov}, \citenamefont
  {Andersen}, \citenamefont {Jullien}, \citenamefont {Babcock}, \citenamefont
  {Chastagnier}, \citenamefont {Chung}, \citenamefont {Humblot}, \citenamefont
  {Lelievre-Berna},\ and\ \citenamefont {Tasset}}]{Pet2006pb}%
  \BibitemOpen
  \bibfield  {author} {\bibinfo {author} {\bibfnamefont {A.~K.}\ \bibnamefont
  {Petoukhov}}, \bibinfo {author} {\bibfnamefont {K.~H.}\ \bibnamefont
  {Andersen}}, \bibinfo {author} {\bibfnamefont {D.}~\bibnamefont {Jullien}},
  \bibinfo {author} {\bibfnamefont {E.}~\bibnamefont {Babcock}}, \bibinfo
  {author} {\bibfnamefont {J.}~\bibnamefont {Chastagnier}}, \bibinfo {author}
  {\bibfnamefont {R.}~\bibnamefont {Chung}}, \bibinfo {author} {\bibfnamefont
  {H.}~\bibnamefont {Humblot}}, \bibinfo {author} {\bibfnamefont
  {E.}~\bibnamefont {Lelievre-Berna}}, \ and\ \bibinfo {author} {\bibfnamefont
  {F.}~\bibnamefont {Tasset}},\ }\bibfield  {title} {\enquote {\bibinfo {title}
  {Recent advances in polarized {H}e-3 spin filters at the {ILL}},}\
  }\href@noop {} {\bibfield  {journal} {\bibinfo  {journal} {Phys. B}\ }\textbf
  {\bibinfo {volume} {385}},\ \bibinfo {pages} {1146} (\bibinfo {year}
  {2006}{\natexlab{b}})}\BibitemShut {NoStop}%
\bibitem [{\citenamefont {Zimmer}\ \emph {et~al.}(1999)\citenamefont {Zimmer},
  \citenamefont {Muller}, \citenamefont {Hautle}, \citenamefont {Heil},\ and\
  \citenamefont {Humblot}}]{Zim1999plb}%
  \BibitemOpen
  \bibfield  {author} {\bibinfo {author} {\bibfnamefont {O.}~\bibnamefont
  {Zimmer}}, \bibinfo {author} {\bibfnamefont {T.~M.}\ \bibnamefont {Muller}},
  \bibinfo {author} {\bibfnamefont {P.}~\bibnamefont {Hautle}}, \bibinfo
  {author} {\bibfnamefont {W.}~\bibnamefont {Heil}}, \ and\ \bibinfo {author}
  {\bibfnamefont {H.}~\bibnamefont {Humblot}},\ }\bibfield  {title} {\enquote
  {\bibinfo {title} {High precision neutron polarization analysis using opaque
  spin filters},}\ }\href@noop {} {\bibfield  {journal} {\bibinfo  {journal}
  {Phys. Lett. B}\ }\textbf {\bibinfo {volume} {455}},\ \bibinfo {pages} {62}
  (\bibinfo {year} {1999})}\BibitemShut {NoStop}%
\bibitem [{\citenamefont {Bigelow}, \citenamefont {Nacher},\ and\ \citenamefont
  {Leduc}(1992)}]{Big1992jp}%
  \BibitemOpen
  \bibfield  {author} {\bibinfo {author} {\bibfnamefont {N.}~\bibnamefont
  {Bigelow}}, \bibinfo {author} {\bibfnamefont {P.}~\bibnamefont {Nacher}}, \
  and\ \bibinfo {author} {\bibfnamefont {M.}~\bibnamefont {Leduc}},\ }\bibfield
   {title} {\enquote {\bibinfo {title} {Accurate optical measurement of nuclear
  polarization in optically pumped {H}e-3 gas},}\ }\href@noop {} {\bibfield
  {journal} {\bibinfo  {journal} {J. Phys. II}\ }\textbf {\bibinfo {volume}
  {2}},\ \bibinfo {pages} {2159} (\bibinfo {year} {1992})}\BibitemShut
  {NoStop}%
\bibitem [{\citenamefont {Mund}\ \emph {et~al.}(2013)\citenamefont {Mund},
  \citenamefont {Maerkisch}, \citenamefont {Deissenroth}, \citenamefont
  {Krempel}, \citenamefont {Schumann}, \citenamefont {Abele}, \citenamefont
  {Petoukhov},\ and\ \citenamefont {Soldner}}]{Mund:2012fq}%
  \BibitemOpen
  \bibfield  {author} {\bibinfo {author} {\bibfnamefont {D.}~\bibnamefont
  {Mund}}, \bibinfo {author} {\bibfnamefont {B.}~\bibnamefont {Maerkisch}},
  \bibinfo {author} {\bibfnamefont {M.}~\bibnamefont {Deissenroth}}, \bibinfo
  {author} {\bibfnamefont {J.}~\bibnamefont {Krempel}}, \bibinfo {author}
  {\bibfnamefont {M.}~\bibnamefont {Schumann}}, \bibinfo {author}
  {\bibfnamefont {H.}~\bibnamefont {Abele}}, \bibinfo {author} {\bibfnamefont
  {A.}~\bibnamefont {Petoukhov}}, \ and\ \bibinfo {author} {\bibfnamefont
  {T.}~\bibnamefont {Soldner}},\ }\bibfield  {title} {\enquote {\bibinfo
  {title} {{Determination of the Weak Axial Vector Coupling from a Measurement
  of the Beta-Asymmetry Parameter {$A$} in Neutron Beta Decay}},}\ }\href
  {\doibase 10.1103/PhysRevLett.110.172502} {\bibfield  {journal} {\bibinfo
  {journal} {Phys. Rev. Lett.}\ }\textbf {\bibinfo {volume} {110}},\ \bibinfo
  {pages} {172502} (\bibinfo {year} {2013})},\ \Eprint
  {http://arxiv.org/abs/1204.0013} {arXiv:1204.0013 [hep-ex]} \BibitemShut
  {NoStop}%
\bibitem [{\citenamefont {Soyama}\ \emph {et~al.}(1995)\citenamefont {Soyama},
  \citenamefont {Suzuki}, \citenamefont {Kodiara}, \citenamefont {Ebisawa},
  \citenamefont {Kawabata},\ and\ \citenamefont {Tasaki}}]{Soy1995pb}%
  \BibitemOpen
  \bibfield  {author} {\bibinfo {author} {\bibfnamefont {K.}~\bibnamefont
  {Soyama}}, \bibinfo {author} {\bibfnamefont {M.}~\bibnamefont {Suzuki}},
  \bibinfo {author} {\bibfnamefont {T.}~\bibnamefont {Kodiara}}, \bibinfo
  {author} {\bibfnamefont {T.}~\bibnamefont {Ebisawa}}, \bibinfo {author}
  {\bibfnamefont {Y.}~\bibnamefont {Kawabata}}, \ and\ \bibinfo {author}
  {\bibfnamefont {S.}~\bibnamefont {Tasaki}},\ }\bibfield  {title} {\enquote
  {\bibinfo {title} {Transmission characteristics of a supermirror bender},}\
  }\href@noop {} {\bibfield  {journal} {\bibinfo  {journal} {Phys. B}\ }\textbf
  {\bibinfo {volume} {213}},\ \bibinfo {pages} {951} (\bibinfo {year}
  {1995})}\BibitemShut {NoStop}%
\bibitem [{\citenamefont {Petukhov}()}]{Pet:un}%
  \BibitemOpen
  \bibfield  {author} {\bibinfo {author} {\bibfnamefont {A.~K.}\ \bibnamefont
  {Petukhov}},\ }\href@noop {} {\bibinfo  {journal} {unpublished}\
  }\BibitemShut {NoStop}%
\end{thebibliography}%

\appendix

\section{Restrictions on the number of plates in a solid-state polarizer of cold neutrons}\label{app:RestrictionOnMirrorNumber}

Consider a stack of double-side coated plates built by setting each plate on top of the previous one. Because of imperfections in the plate geometry (a plate may be twisted, for example), or because of small dust particles sitting between the plates, the neighbouring plates may be not perfectly parallel. If the previous plate with index $i-1$ has orientation $\boldsymbol{n}_{i-1}$, plate $i$ has orientation $\boldsymbol{n}_i$:  
\begin{equation}
\boldsymbol{n}_i=\boldsymbol{n}_{i-1}+\xi_i.
\label{eq:A1}
\end{equation}
Here, $\boldsymbol{n}_i$ and $\boldsymbol{n}_{i-1}$ are unit vectors normal to the plates' surfaces and $\xi_i$ is a vector representing a possible error in positioning. For simplicity, we restrict ourselves to one dimension, see Fig.~\ref{fig:Scheme}:
\begin{equation}
\theta_i=\theta_{i-1}+\xi_i.
\label{eq:A2}
\end{equation}

The sequence Eq.~(\ref{eq:A2}) may be treated as a Random Walk with a continuous Gaussian random step $\xi_i$:
\begin{equation}
\boldsymbol{E}[\xi_i]=0,~~~\boldsymbol{E}[\xi_i \xi_j]=\sigma^2\delta_{i,j},
\label{eq:A3}
\end{equation}
where $E[x]$ is the expectation value of $x$ and $\delta$ is the Kronecker symbol.

Eq.~(\ref{eq:A2}) may be also written as follows:
\begin{equation}
\theta_i=\theta_{i-1}+\xi_i=\theta_{i-2}+\xi_{i}+\xi_{i-1}=\dots=\theta_1+\sum_{j=2}^i \xi_j.
\label{eq:A4}
\end{equation}

Considering the orientation of the first plate in the stack as reference, we obtain:
\begin{eqnarray}
\boldsymbol{E}[\theta_i-\theta_1]&=&0,\nonumber\\
\boldsymbol{E}[(\theta_i-\theta_1)(\theta_i-\theta_1)]&=&\boldsymbol{E}[(\sum_{k=2}^i{\xi_k})(\sum_{m=2}^i{\xi_m})] \nonumber\\
&=&\sum_{k=2}^i\sum_{m=2}^i\boldsymbol{E}[\xi_k\xi_m] \nonumber\\
&=&(i-1)\sigma^2,
\label{eq:A5}
\end{eqnarray}
for $i>1$.

Eq.~(\ref{eq:A5}) tells that the variance of the error in the positioning of plate $i$ grows linearly with index $i$.

Applying this model to the solid state polarizer, we associate $\sigma$ with the one-step precision in mounting of a single plate. We also assume a $100\%$ reflectivity for neutrons incident on the stack of plates within the incident angle $\alpha=\pm\theta_c$, where $\theta_c$ is the critical angle for the reflecting coating:
\begin{equation}
\theta_c=17.3m\lambda.
\label{eq:A7}
\end{equation}
Here, $\theta_c$ is given in $mrad$, $m$ is the index of the SM coating, and the neutron wavelength $\lambda$ is given in $nm$.

Consider neutrons hitting the plate from the edge as shown in Fig.~\ref{fig:InterStack} and propagating through the substrate between two reflecting planes. Even for a perfectly collimated beam, some neutrons would hit the coating due to the angular dispersion in plate positioning. 

Depending on the hitting angle, neutrons would be reflected for $|\theta|<\theta_c$, or penetrate through the coating and be absorbed in the Gd absorbing layers. The dispersion of hitting angles follows the dispersion of angular positioning of the plates. Therefore, the dispersion of angular orientation of individual plates in a solid state polarizer leads to a decrease in the device transmission (valid for a polarizer, but also for a collimator or deflector) as well as to a decrease of polarization of the transmitted beam due to the opening of additional channels for spin leakage.

In practice, the direction of the incident beam is not aligned to the very first assembled plate and the beam illuminates all plates in the stack, therefore, instead of the variance Eq.~(\ref{eq:A5}) valid for the mirrors on plate with index $i$ we have to consider the variance of the hitting angle $\theta$ averaged over all plates in the stack:
\begin{equation}
\delta_{\rm stack}^2={\langle (\theta_i)^2\rangle}_{\rm stack}=\frac{1}{n-1}\sum_{i=2}^n{(i-1)\sigma^2}=\frac{1}{2}n\sigma^2.
\label{eq:A8}
\end{equation}
Here, $n$ is the number of plates in the stack. 

To provide a condition that neutrons hitting the wall would have a chance to survive and continue to propagate with the probability of $0.95$, one needs to limit the number of plates in the stack such that $\theta_c>2\delta_{\rm stack}$, or:
\begin{equation}
n<n_c\approx\frac{1}{2}\left( \frac{\theta_c}{\sigma} \right)^2.
\label{eq:A9}
\end{equation}

For a stack inclined by an angle $\alpha$ relative to the incident beam, the constrain Eq.~(\ref{eq:A9}) has to be modified by replacing $\theta_c$ by $\theta_c-\alpha$.

In the above example, we consider a perfectly collimated incident beam. For a beam with a finite angular divergence $\delta_{\rm beam}$, the angular variance of the plates in the stack increases the variance of hitting angles which may be translated into an ``effective'' increase of the incident beam divergence:
\begin{equation}
\delta_{\rm eff}=\sqrt {\delta_{\rm beam}^2+\delta_{\rm stack}^2}.
\label{eq:A10}
\end{equation}

\end{document}